\documentclass[11pt,a4paper]{article}
\usepackage{jcappub}
\usepackage[a4paper,top=1in,bottom=1in,left=15mm,right=15mm,footskip=0.25in]{geometry}

\usepackage{cleveref}[2012/02/15]
\crefformat{footnote}{#2\footnotemark[#1]#3}

\usepackage{doi}

\usepackage[flushleft]{threeparttable}

\usepackage{amsmath}
\usepackage{amsbsy}
\usepackage{mathtools}
\usepackage{float}
\usepackage{datetime}
\usepackage{textpos}
\usepackage{booktabs}
\usepackage{graphicx}
\usepackage{mathrsfs}
\usepackage{adjustbox}
\usepackage{epsfig}
\usepackage{color}
\usepackage{bm}
\usepackage{graphicx}
\usepackage{physics}
  
\usepackage[all]{hypcap}

\usepackage{natbib}

\providecommand{\adsurl}[1]{\href{#1}{ADS}}
 
\providecommand{\url}[1]{\href{#1}{#1}}
\usepackage{url}

\usepackage{accents}

\usepackage{amsmath}

\usepackage{graphicx}
\usepackage{amsmath}
\usepackage{float}

\usepackage{times}
\usepackage{mathptmx}

\usepackage{hhline}

\providecommand{\abs}[1]{\lvert#1\rvert}

\usepackage{subfigure}

\usepackage{placeins}
\usepackage{chngcntr}

\DeclareMathAlphabet\mathbfcal{OMS}{cmsy}{b}{n}

\def\alt{\raise0.3ex\hbox{$\;<$\kern-0.75em\raise-1.1ex\hbox{$\sim\;$}}}
\def\agt{\raise0.3ex\hbox{$\;>$\kern-0.75em\raise-1.1ex\hbox{$\sim\;$}}}

\newcommand{\bw}{\begin{widetext}}
\newcommand{\ew}{\end{widetext}}

\newcommand{\lsim}{\,\rlap{\raise 0.35ex\hbox{$<$}}{\lower 0.7ex\hbox{$\sim$}}\,}
\newcommand{\gsim}{\,\rlap{\raise 0.35ex\hbox{$>$}}{\lower 0.7ex\hbox{$\sim$}}\,}

\def\lesssim{\mathrel{\hbox{\rlap{\hbox{\lower3pt\hbox{$\sim$}}}\hbox{\raise2pt\hbox{$<$}}}}}
\def\gtrsim{\mathrel{\hbox{\rlap{\hbox{\lower3pt\hbox{$\sim$}}}\hbox{\raise2pt\hbox{$>$}}}}}

\def\xlinkspace#1 #2{%
 \ifx\relax#2%
 \xlinkdash#1-\relax
 \else
 \xlinkdash#1 -\relax
 \expandafter\xlinkspace\expandafter#2%
 \fi}

\def\xlinkdash#1-#2{%
 \ifx\relax#2%
 \tmp{#1}%
 \else
 \tmp{#1-}%
 \expandafter\xlinkdash\expandafter#2%
 \fi}

\newcommand{\newtext}[1]{\textcolor{black}{#1}}

\usepackage[normalem]{ulem}

\title{The effect of kick velocities on the spatial distribution of millisecond pulsars and implications for the Galactic center excess}
\author{Harrison Ploeg and Chris Gordon}
\affiliation{School of Physical and Chemical Sciences, University of Canterbury, Christchurch, New Zealand}
\emailAdd{hzp10@uclive.ac.nz}
\emailAdd{chris.gordon@canterbury.ac.nz}

\abstract{Recently it has become apparent that the Galactic center excess (GCE) is spatially correlated with the stellar distribution in the Galactic bulge. This has given extra motivation for the unresolved population of millisecond pulsars (MSPs) explanation for the GCE. However, in the ``recycling'' channel the neutron star forms from a core collapse supernovae that undergoes a random ``kick'' due to the asymmetry of the explosion. This would imply a smoothing out of the spatial distribution of the MSPs. We use $N$-body simulations to model how the MSP spatial distribution changes.  We estimate the probability distribution of natal kick velocities
using the resolved gamma-ray MSP proper motions, where MSPs have random velocities relative to the circular motion with a scale parameter of  $77\pm6$~km/s. We find that, due to the natal kicks,  there is an approximately 10\% increase in each of the bulge MSP spatial distribution dimensions and also the bulge MSP distribution becomes less boxy but is still far from being spherical.  }

\begin{document}

\maketitle

\section{Introduction}

The Galactic Center Excess (GCE) is an extended source of gamma radiation in the central region of the Galaxy found in the Fermi Large Area Telescope (Fermi-LAT) data. When first discovered, its apparently spherically symmetric profile and spectrum peaking at a few GeV suggested that it may be evidence of self-annihilating weakly interacting massive particles (WIMPs) distributed according to a Navarro-Frenk-White (NFW) profile \cite{Goodenough:2009gk,Hooper_2011,Abazajian:2012pn,Gordon:2013vta}. An alternative scenario that was proposed was one in which the GCE was produced by a population of unresolved millisecond pulsars (MSPs) \cite{Abazajian:2010zy}. More recently, evidence has suggested that in fact the GCE is not spherically symmetric but actually is correlated with the stellar mass in the Galactic bulge and so has a more boxy morphology \cite{Macias_2018,Bartels2017, Macias19, Abazajian2020,Coleman19,Calore2021}.
Although one recent study, using different methods, still argues for a spherically symmetric GCE \cite{DiMauro2021}. If the GCE does trace the stellar mass of the bulge this would favor the  MSP (or other dim unresolved astrophysical point source) explanation.
Additionally, several studies have claimed to find a non-Poissonian component to the GCE \cite{Bartels:2015aea,Lee:2015fea}
which may be further evidence for the MSP explanation. However, there is some controversy regarding the level of systematics in this approach \citep[e.g.,][]{LeaneSlatyer2019,LeaneSlatyer2020,
Leane2020,Buschmann2020,Chang2020,Calore2021}.

In the ``recycling'' model of MSP formation a neutron star is spun up to millisecond periods through the transfer of mass from a binary companion. This requires that the binary system survives the kick produced by any asymmetry in the core collapse supernova explosion \cite{Bhattacharya1991}.
However, an alternative to the recycling channel is 
accretion induced collapse of white dwarfs into neutron stars which may produce more than half of all observed MSPs \citep{Ferrario2007,Hurley2010,Ruiter2019,Gautam2021}. In this case the system does not receive a significant natal kick \citep{Fryer1999,Kitaura2006}. 
This would imply that the MSPs have much smaller peculiar velocities 
in comparison to the recycling model \citep{Lyne1994, Wongwathanarat2013, Bear2018}.

Ploeg et al.~\cite{Ploeg2020} (hereafter referred to as P20) modelled Fermi-LAT detected MSPs as having a Maxwell distributed peculiar velocity with the scale parameter $\sigma_v$ found to be $77 \pm 6$ km s$^{-1}$ where we quote error bars at the 68\% confidence interval throughout this article. This velocity applies for disk MSPs and we assume it will not be significantly different for bulge MSPs. Although the star formation histories are very different in the bulge and disk, as can be seen from Fig.~6 of P20 the probability distribution of luminosities in the bulge and disk only differ by a few percent. 
Also, as can be seen from Fig.~11 of P20, the bulge and disk have a ratio of number of MSPs formed per solar mass which is within one error bar of each other.
 Based on this, we assume that the bulge and disk have the same mix of MSP formation channels and thus the same probability distribution of natal kick velocities.

If the source of the GCE is a population of unresolved MSPs, then the spatial distribution may be smoothed to some degree relative to the stellar mass in the bulge. Eckner et al.~\cite{Eckner2018} used the  virial theorem to estimate the ``smoothing length" of MSPs as $700$ -- $900$ pc for kicks $\lesssim 70$ km s$^{-1}$. However, they assumed a spherically symmetric spatial distribution for the MSPs. 

In this article we use $N$-body simulations to estimate what are the effects of MSP kicks for a boxy bulge like distribution. In Section~\ref{method} we explain our method. Our results are given in Section~\ref{results}, and our conclusions in Section~\ref{conclusion}.

\section{Method}
\label{method}
For this work we use the code of Bedorf et al.~\cite{Bedorf2012} to run $N$-body simulations in order to model the Milky Way.\footnote{Available at: https://github.com/treecode/Bonsai} We use parameters corresponding to initial conditions labelled MWa, MWb and MWc0.8 as denoted by Fujii et al.~\cite{Fujii2019} as they were the initial conditions that led to the best fitting simulations to Galaxy observations that Fujii et al.\ found.
\newtext{The initial conditions consist of a parametric description of the dark matter halo, Galactic disk and bulge.}
Comparing to bulge kinematics, bar length, and pattern speed observations  Fujii et al. \ found
\newtext{MWa was the initial conditions which gave the best fit to the data, followed by MWb, and then MWc0.8.}
For each initial condition model we generated a total of $30$ million disk, bulge and dark matter halo particles. These initial populations are generated using the methods of Kuijken and Dubinski \cite{Kuijken1995}, Widrow and Dubinski \cite{Widrow2005}, and Widrow et al.~\cite{Widrow2008}.\footnote{We used the implementation at: https://github.com/treecode/Galactics.parallel}
As in Fujii et al.~\cite{Fujii2019}, we use time-steps of $\sim0.6$ Myr, an opening angle of $0.4$ and ran the simulation for $10$ Gyr. However, we use a softening length of $30$ pc. Also, our dark-matter halo particles have a mass $8$ times larger than the disk and bulge particles. 
Taking into account the masses of the various components, this implies that,
out of the 30 million particles,  approximately $10$ million particles represent  stellar mass and the remainder represent dark matter. 

In order to model the density of MSPs, we additionally include massless (so they do not affect the simulation) disk and bulge particles that are given a normally distributed perturbation to each component of their velocity vector with mean zero and standard deviation $\sigma_k$. The kick velocity magnitude is therefore Maxwell distributed. The probability density function of a Maxwell distribution can be written as:
\begin{equation}
    p(x) = \sqrt{\frac{2}{\pi}} \frac{x^2 \exp\left(-x^2/2\sigma^2\right)}{\sigma^3}
    \label{eq:Maxwell}
\end{equation}
\noindent where $x$ is the magnitude of a three dimensional vector with components sampled from the normal distribution $\mathcal{N}(0,\sigma^2)$. For each model we try a case where the kicks occurred at the beginning of the $N$-body simulations and a case where the kicks occur randomly with a uniform rate over the course of the $10$ Gyr. \newtext{These cases approximate scenarios in which MSPs are largely born early or at a relatively steady rate. From these two extremes, we can estimate the sensitivity of our results to the MSP age distribution. }

The first step is to estimate the kick velocity scale required to produce a peculiar velocity distribution consistent with P20 where for the best model the peculiar velocity scale parameter was $\sigma_v=77 \pm 6$ km s$^{-1}$. We do this by running each initial condition model with $41$ populations of $10^5$ kicked particles with $\sigma_k$ between $70$ and $110$ km s$^{-1}$. We separate the velocity of each particle into two components:
\begin{equation}
    \boldsymbol{v} = \boldsymbol{v}_c + \boldsymbol{v}_p
\end{equation}
\noindent where $\boldsymbol{v}_c$ is the velocity of a particle on a circular orbit around the center of the galaxy and $\boldsymbol{v}_p$ is the peculiar velocity. The magnitude of $\boldsymbol{v}_c$ for a particle with coordinates $x$, $y$, and $z$ can be evaluated using the centripetal force: 
\begin{equation}
    \lVert \boldsymbol{v}_c \rVert = \sqrt{\lVert\boldsymbol{a}_c(x,y,0)\rVert R}
\end{equation}
\noindent where $\boldsymbol{a}_c(x,y,z)$ is the acceleration toward the center of the galaxy which can be obtained from a small modification to the the $N$-body code
and $R^2=x^2+y^2$. For $R$ outside the bulge region, and for small peculiar velocity, we are therefore assuming that particles are rotating with the disk, with $\boldsymbol{v}_c$ the rotation velocity of the disk at $R$ \cite{Verbunt2017}.

We use a maximum likelihood estimate of the final $\sigma_v$ for each initial $\sigma_k$. For a set of $N$ particles with peculiar velocities $v_1$, ..., $v_N$, the log-likelihood is obtained by assuming velocities have a Maxwell distribution:
\begin{equation}
    \log(L) = \frac{N}{2} \log(\frac{2}{\pi}) - 3 N \log(\sigma_v) + \sum_{i=1}^N 2 \log(v_i) - \frac{v_i^2}{2 \sigma_v^2}
\end{equation}
\noindent and therefore
\begin{equation}
    \frac{\dd\log(L)}{\dd\sigma_v} = -\frac{3 N}{\sigma_v} + \sum_{i=1}^N \frac{v_i^2}{\sigma_v^3} \, .
\end{equation}
\noindent Then solving for $\sigma_v$ where $\dd\log(L)/\dd\sigma_v = 0$, we find the maximum likelihood estimate for $\sigma_v$ is:
\begin{equation}
    \hat{\sigma_v} = \sqrt{\frac{\sum_{i=1}^N v_i^2}{3 N}}
\end{equation}
This is done for particles where $4 \textrm{ kpc} \leq R \leq 12 \textrm{ kpc}$ and $\lvert z \rvert \leq 2 \textrm{ kpc}$, ensuring we are estimating the peculiar velocity distribution scale parameter for particles in the disk region from which gamma-ray MSPs are most likely to be resolved and where $\boldsymbol{v}_c$ approximately represents disk rotation.

  \newtext{Once we have a best fitting $\sigma_k$ we use that $\sigma_k$ to generate an additional population of kicked particles. We rerun the $N$-body simulations for each set of initial conditions (MWa, MWb, MWc0.8) including in each initial condition case two new populations of massless particles where each new population has $2\times10^6$ particles. These new populations are made massless so as not to influence the gravitational potential of the $N$-body simulation. The first  massless populations is  sampled from the initial condition model. For the second population we duplicate the first population and then add a Maxwell distributed kick with scale parameter $\sigma_k$ at the beginning of the simulation to each particle in the second population. The simulations are then run for 10 Gy to see what the final state of the two massless populations is for each set of initial conditions.
We then repeat this process but instead of the second population having a Maxwell distributed kick added at the beginning of the simulation we add it randomly to members of the second population at a uniform rate throughout the simulation. So in the end for each initial condition we have two massless populations of unkicked particles, one massless population of particles kicked at the beginning and one for particles kicked at a uniform rate throughout the simulation. We used two populations of massless unkicked particles to make the kicked at beginning cases and uniform kick rate cases as independent as possible from each other.}

\newtext{In order to understand the effect of MSP kicks on the final spatial distribution of MSPs we found it clearer to have  a parametric description of the spatial distribution. Then by comparing the parameters for the kick and no-kick cases we can have a succinct way of characterising the effects of the kicks.}
We use Markov Chain Monte Carlo (MCMC) to fit the parametric model to both final particle distributions. 
\newtext{The form of our parametric model for the final spatial distribution of MSPs was chosen based on previous studies and also such that there were not excessive  residuals between the model fits and the simulations.} 
Our model of the final distributions consists of four components: a spherically symmetric bulge, a potentially non-spherical bulge also known as a bar, a long bar and a disk. The spherically symmetric bulge component uses the Hernquist model \cite{Hernquist1990}:
\begin{equation}
    \rho_{\rm Hernquist} (r) \propto \frac{1}{\left(r/a_b \right) \left(1 + r/a_b\right)^3}
\end{equation}
\noindent where $r^2 = x^2 + y^2 + z^2$ and $a_b$ is a free parameter. Our  initial conditions models, based on those of  Fujii et al.~\cite{Fujii2019}, also include a component distributed according the Hernquist model
\newtext{which would naturally have different parameter values  to the one we fitted after 10~Gy of evolution.}
Our bar model is distributed as:
\begin{equation}
\rho_{\rm bar} (R_s) \propto K_0(R_s)
\times \begin{cases}
      1 & R \leq R_{\rm end} \\
      \exp(-(R - R_{\rm end})^2/h^2_{\rm end}) & R > R_{\rm end} \\
\end{cases}
\label{eq:bar}
\end{equation}
\noindent where $K_0$ is the modified Bessel function of the second kind and where:
\newtext{
\begin{equation}
R_s = \left(R_\perp^{C_\parallel} + \left(\frac{\abs{z}}{z_b} \right)^{C_\parallel}\right)^{1/C_\parallel}
\end{equation}
\begin{equation}
R_\perp = \left(\left(\frac{\abs{x}}{x_b} \right)^{C_\perp} + \left(\frac{\abs{y}}{y_b} \right)^{C_\perp}\right)^{1/C_\perp}\,.
\end{equation}
}
\noindent The free parameters are $C_\parallel$, $C_\perp$, $x_b$, $y_b$, $z_b$ and $R_{\rm end}$ with $h_{\rm end}$ fixed at $\sqrt{\frac{1}{2}}$ kpc. 
The effective radius is $R_s$; the scale lengths are $x_b$, $y_b$, and $z_b$; and $C_\perp$ and $C_{\parallel}$ are the face-on
and edge-on shape parameters.
The bar shape is elliptical in the corresponding direction when $C_\perp,C_{\parallel} = 2$, diamond-shaped when $C_\perp,C_{\parallel} < 2$, and boxy when $C_\perp,C_{\parallel} > 2$.
The Gaussian function with
scale length $h_{\rm end}$ in Eq. (\ref{eq:bar}) 
 truncates the bar at radius $R_{\rm end}$.
The modified Bessel function was also used in Cao et al.~\cite{Cao:2013dwa} to model the distribution of red clump giants, but with no cutoff and with $C_\parallel = 4$ and $C_\perp = 2$. For the long bar we use \cite{Wegg2015}:
\begin{equation}
\label{eq:long_bar}
\rho_{\rm long~bar} (x, y, z) \propto \exp\left( -\left(\left(\frac{\abs{x}}{x_{\rm lb}} \right)^{C_{\perp,\rm lb}} + \left(\frac{\abs{y}}{y_{\rm lb}} \right)^{C_{\perp,\rm lb}}\right)^{1/C_{\perp,\rm lb}} \right) \exp\left(-\frac{\abs{z}}{z_{\rm lb}} \right) \textrm{Cut}\left(\frac{R - R_{\rm out}}{\sigma_{\rm out}} \right) \textrm{Cut}\left(\frac{R_{\rm in} - R}{\sigma_{\rm in}} \right)
\end{equation}
\noindent where $C_{\perp,\rm lb}$, $x_{\rm lb}$, $y_{\rm lb}$, $z_{\rm lb}$, $R_{\rm out}$ and $R_{\rm in}$ are free parameters, $\sigma_{\rm out} = \sigma_{\rm in} = \sqrt{\frac{1}{2}}$ kpc and:
\begin{equation}
    \textrm{Cut}\left(x\right) = \begin{cases}
      \exp(-x^2) & x > 0 \\
      1 & x \leq 0 \\
\end{cases}
\end{equation}
\noindent \newtext{Note that Fujii et. al~\cite{Fujii2019} did not need to include any bar components in their initial conditions as the bar components evolve naturally over the 10~Gy from their Hernquist bulge, disk and  halo model initial conditions.}

Finally, we have a disk with a central hole:
\begin{equation}
    \rho_{\rm disk} (x, y, z) \propto \exp(-R^2/2\sigma_r^2) \exp(-\abs{z}/z_0) H(x,y)
\end{equation}
\noindent where
$\sigma_r$ and $z_0$ are free parameters and for the hole we use the form adopted by Freudenreich \cite{Freudenreich:1997bx}:
\begin{equation}
    H(x,y) = 1 - \exp\left(-\left(R_H/O_R\right)^{O_N} \right)
\end{equation}
with:
\begin{equation}
    R_H^2 = (x)^2 + (\epsilon y)^2
\end{equation}
\noindent where $\epsilon$, $O_R$ and $O_N$ are also free parameters. \newtext{Note that Fujii et. al \cite{Fujii2019} did not have a hole in their initial disk model. This hole evolves naturally through the process of the formation of the bar over their 10 Gy simulation.}

The total number of particles in our simulations are fixed. So we do not have to include the number of particles as part of our likelihood. Therefore the probability of having an $N$-body particle at position ${x,y,z}$ will be proportional to the density of our model ($\rho$) at ${x,y,z}$.
We have for each component of the model a parameter giving the probability a particle is from that component. We treat the probability of an $N$-body particle being from a component of the density distribution as parameters. These parameters, $P({\rm Disk})$, $P({\rm Bar})$, $P({\rm Hernquist})$ and $P({\rm Long~Bar})$, have a Dirichlet prior \cite{Betancourt2013}. This prior constrains
$$
 P({\rm Disk})  + P({\rm Bar}) + P({\rm Hernquist})  + P({\rm Long~Bar})=1
$$
and is uniformly distributed over any values of these parameter satisfying this condition.
The log-likelihood is then:
\begin{equation}
    \log(L) = \sum_i^N \log(\rho(x_i, y_i, z_i))
\end{equation}
\noindent where $x_i$, $y_i$ and $z_i$ are the coordinates of a particle, $N$ is the number of particles, and $\rho$ is the density of the model:
\begin{equation}
    \rho(x, y, z) = P({\rm Disk}) \rho_{\rm disk}(x, y, z) + P({\rm Bar}) \rho_{\rm bar}(x, y, z) + P({\rm Hernquist}) \rho_{\rm Hernquist}(x, y, z) + P({\rm Long~Bar}) \rho_{\rm long~bar}(x, y, z)
\end{equation}
All scale parameters are given a prior so they are uniform in $\log(\theta)$ where $\theta\in \{a_b, x_b, y_b, z_b, x_{\rm lb}, y_{\rm lb}, z_{\rm lb},\sigma_r,z_0, O_R\}$ and this implies $p(\theta) \propto 1/\theta$.
In calculating the likelihood, we do not include particles for which $R>12$ kpc or $\abs{z} > 3$. It can be seen in Fujii et al.~\cite{Fujii2019} that the scale height of the disk may start to decline between $10\lesssim R \lesssim 15$~kpc. We also don't want the fit to be affected by particles that may have been kicked well out of the galaxy. 
The likelihood ($L$) is insensitive to being multiplied by a constant but that constant has to be the same for all parameters of our combined model.
To accommodate this we normalize each density component such that
$$
\int_{R \leq 12~{\rm kpc},\abs{z} \leq 3~{\rm kpc}} \rho_i(x,y,z)\, {\rm d}x\,{\rm d}y\,{\rm d}z=1
$$
where $i\in$~\{disk, bar, Hernquist, long bar\}.
This integral is estimated with importance sampling.
We use a set of random numbers which are transformed into the points at which we evaluate the density models in order to estimate the normalization constant. In order to stabilize the estimation of the likelihood function these numbers are always the same every time we perform the importance sampling within a particular chain.

After running the $N$-body simulations, we shift the coordinates of the particles so that the center of mass is at the origin, then rotate so the bar is along the $x$-axis. The bar angle is estimated using the method described in Fujii et al.~\cite{Fujii2019}. However, we add four parameters that we expect to be near zero to allow a further shift in the center and clockwise rotation of the model. These are $\alpha$, $x_{\rm center}$, $y_{\rm center}$ and $z_{\rm center}$, with the latter three parameters in parsecs, so:
\begin{equation}
  \begin{aligned}
    x_{\rm data} &= \cos(\alpha) x + \sin(\alpha) y + x_{\rm center} / 1000 \\
    y_{\rm data} &= -\sin(\alpha) x + \cos(\alpha) y + y_{\rm center} / 1000 \\
    z_{\rm data} &= z + z_{\rm center} / 1000
  \end{aligned}
\end{equation}
\noindent where $x_{\rm data}$, $y_{\rm data}$ and $z_{\rm data}$ are coordinates in the coordinate system of the $N$-body simulation. In estimating the peculiar velocity distribution scale parameter, $\hat{\sigma_v}$, above, we assumed $x \approx x_{\rm data}$, $y \approx y_{\rm data}$ and $z \approx z_{\rm data}$.

\newtext{To compare our model fits to the final $N$-body simulations we sample from the model directly, then we bin the samples. We used a simple MCMC to sample from the bar and long bar. Since the disk (excluding the hole) can be turned into an invertible cumulative distributions in $R$ and $z$, and the Hernquist bulge in $r$, we used inverse transform sampling for those. The disk hole was handled by accepting the disk samples with probability $H(x,y)$, i.e. we used rejection sampling for the disk.
}

Our MCMC algorithm used to fit the models is similar to that of Foreman-Mackay et al.~\cite{ForemanMackey2013} with a mixture of the Differential Evolution \cite{TerBraak2006} and snooker updates \cite{TerBraak2008}. We use a simple annealing method in which we divide the log-likelihood by a temperature $T$ which is gradually reduced to $1$. This occurs during the first half of each Markov chain, which we discard. See appendix \ref{app:mcmc} for more details.

\section{Results}
\label{results}

\begin{figure}
    \centering
    \includegraphics[width=0.8\linewidth]{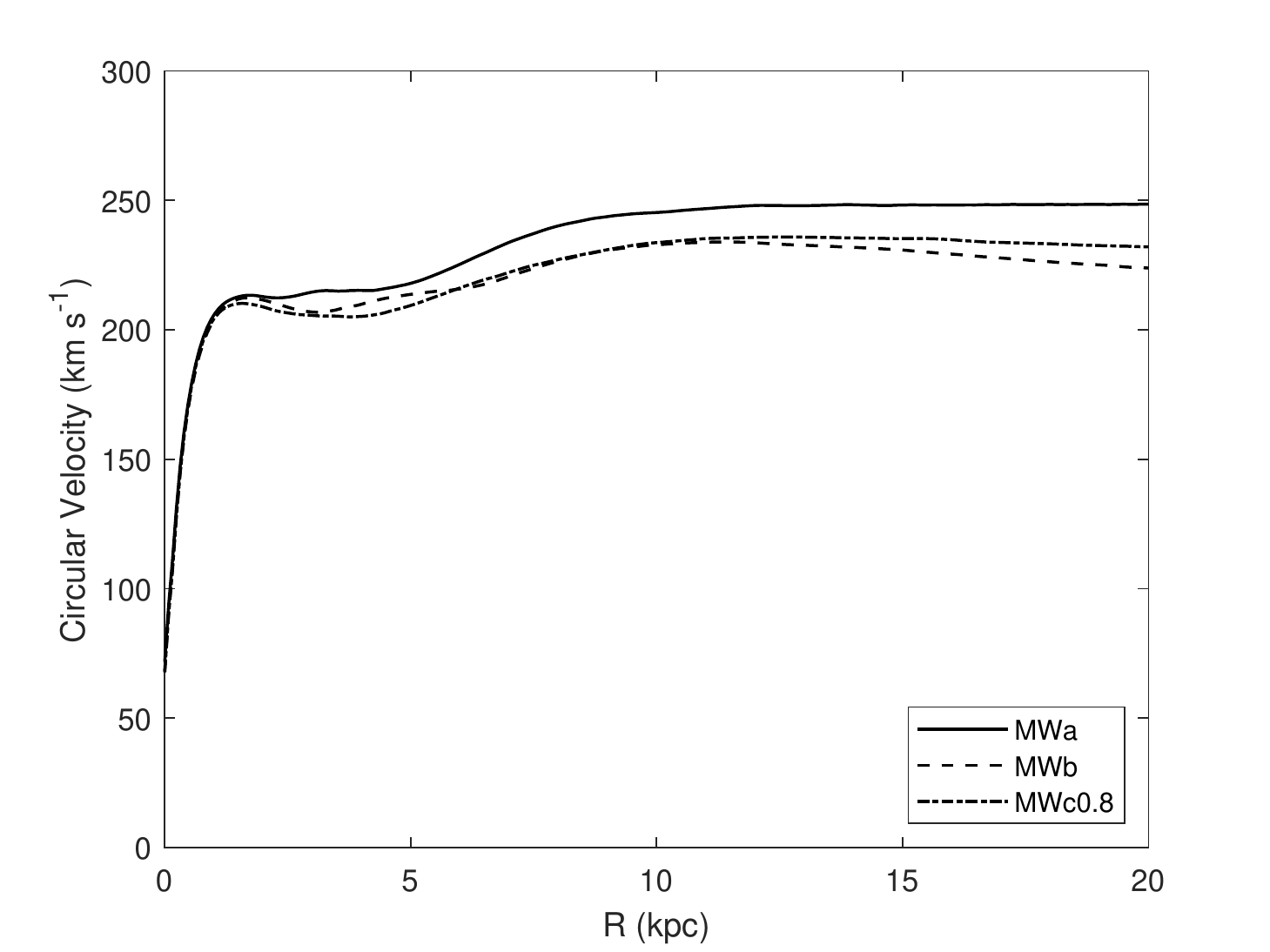}
    \caption{The rotation curves at $t = 10$ Gyr for the three initial condition models  MWa, MWb and MWc0.8. They are consistent with the local circular velocity of the Sun which is measured to be $238 \pm 15$ km s$^{-1}$ \cite{Bland-Hawthorn2016}. The distance between the Sun and the Galactic Center is approximately $8$ kpc. }
    \label{fig:rotation_curves}
\end{figure}

In Table~\ref{tab:kick_velocity} we present the kick velocities $\sigma_k$ that produce peculiar velocity distributions close to $\sigma_v = 77 \pm 6$ km s$^{-1}$ as estimated by P20. We display the rotation curves at $t = 10$ Gyr for the three initial condition models (MWa, MWb and MWc0.8) in Fig.~\ref{fig:rotation_curves}. The central values in Table~\ref{tab:kick_velocity} were used to run $N$-body simulations with a larger number of particles to which we fitted a parametric model. The fitted parameters are shown in Tables~\ref{tab:MWa_params}, \ref{tab:MWb_params} and \ref{tab:MWc0.8_params}. There are three potential sources of uncertainty in the model parameters: the posterior density function, variation in the likelihood between chains as a result of the importance sampling method used to estimate the normalization constant for each model component, and the possibility that Markov chains may get stuck in different local likelihood maxima. We found that for most parameters the posterior distributions overlapped significantly or, in many cases, were indistinguishable. For a few parameters we had outlier chains; these can be seen in the tables as parameters with large, highly asymmetric uncertainties. For the kicked distributions, the disk model hole parameters were very uncertain with different Markov chains settling on a wide range of different values. As this hole tended to be smaller and/or less sharp, we simply removed it, in the kicked cases, by setting $H(x,y) = 1$. We found the long bar model often acts like a second disk component when fitting to the kicked particles, this disk-like long bar was still allowed a hole through the parameter $R_{\rm in}$.

We ran the $N$-body simulations for the two kick rate scenarios separately, but we combine the two sets of Markov chains for the ``No Kick" columns in the parameter tables. Typically, the posterior distributions were very similar for Markov chains generated using these two sets of data, slightly expanding our $68\%$ intervals. 
\newtext{However, we found a difference of order 10 pc in the location of the center between simulations.}
 For this reason, we only report the change in parameters $\alpha$, $x_{\rm center}$, $y_{\rm center}$ and $z_{\rm center}$ for the fits to the kicked distributions by subtracting off the median of the corresponding Markov chains.
For $\alpha$ the fitted $68\%$ intervals for no kick were  $-2^\circ$ to $-1^\circ$ for all models, $\abs{x_{\rm center}}$ and $\abs{y_{\rm center}}$ were always $\lesssim 40$ pc, and $\abs{z_{\rm center}}$ was $< 2$ pc.
\newtext{We are not concerned about this difference as it is negligible compared to the bar extent. }

We show the final density maps generated from the MWa initial conditions in Fig.~\ref{fig:MWa_density_data_fit}. Those for MWb and MWc0.8 were similar and are given in the appendix as Figs.~\ref{fig:MWb_density_data_fit} and \ref{fig:MWc0.8_density_data_fit} respectively. These figures include both the $N$-body particle data as well as the fitted models. These density maps were produced by binning particles (either the $N$-body particles or particles drawn from the fitted model) within $0.25$ kpc of the $x$-$y$, $x$-$z$ and $y$-$z$ planes. The density maps for the fitted models were generated by taking the average in each bin for simulated data generated using 500 random parameter sets from our Markov chains in each case. In Fig.~\ref{fig:MWa_1d_profiles_kick_at_beginning}
we show the $N$-body simulation data and simulated particle distributions along the $x$, $y$ and $z$ axes for the MWa case with the kicks occurring at the beginning.
\newtext{For the model we show the predicted mean number of particles in each bin and also the standard deviation of that prediction. This can be used to test the model fit  as it provides the 68\% posterior predictive intervals \cite{Gelman2013}.}
The corresponding uniform kick rate, MWb, and MWc0.8 cases are displayed in the appendix as Figs.~\ref{fig:MWa_1d_profiles_uniform_kick_rate}, \ref{fig:MWb_1d_profiles_kick_at_beginning}, \ref{fig:MWb_1d_profiles_uniform_kick_rate}, \ref{fig:MWc0.8_1d_profiles_kick_at_beginning} and \ref{fig:MWc0.8_1d_profiles_uniform_kick_rate}.  In these figures, we bin all particles within $0.25$ kpc in the two perpendicular axes.
We display in Fig.~\ref{fig:MWa_data_los} the integrated flux along lines of sight in the central $50^\circ \times 50^\circ$ of the galaxy. The Sun is placed at a distance of $7.9$ kpc, at an angle relative to the bar of $20^\circ$ and at a height of $15$ pc \cite{Coleman19}. This figure was generated by binning particles in Galactic latitude and longitude with weights of $1/d^2$ where $d$ is the distance of a particle from the Sun. We excluded particles within $1$ kpc of the Sun to reduce noise. The corresponding figures for MWb and MWc0.8 are display in the appendix as Figs.~\ref{fig:MWb_data_los} and \ref{fig:MWc0.8_data_los} respectively. 

\begin{table}
\begin{center}
    \begin{tabular}{r||c|c}
          & Kick At Beginning & Uniform Kick Rate \\ \hline \hline
         
         MWa & $93\pm10$ & $85\pm9$ \\ \hline
         MWb & $97\pm10$ & $84\pm10$ \\ \hline
         MWc0.8 & $95\pm10$ & $84\pm9$ \\ \hline
    \end{tabular}
    \label{tab:kick_velocity}
    \caption{ Kick velocity Maxwell distribution (Eq. (\ref{eq:Maxwell})) parameters ($\sigma_k$ in km/s)  that  produce a peculiar velocity distribution where $\sigma_v = 77\pm6$~km/s. }
    \end{center}
\end{table}

\begin{table}
\begin{center}
    \begin{tabular}{r||c|c|c}
         Parameter & No Kick & Kick At Beginning & Uniform Kick Rate \\ \hline \hline
         
         $P({\rm Disk})$ & $0.499\substack{+0.006 \\ -0.005}$ & $0.418\substack{+0.006 \\ -0.006}$ & $0.365\substack{+0.006 \\ -0.004}$ \\ \hline
         $P({\rm Bar})$ & $0.354\substack{+0.004 \\ -0.005}$ & $0.3220\substack{+0.0019 \\ -0.0021}$ & $0.364\substack{+0.005 \\ -0.007}$ \\ \hline
         $P({\rm Hernquist})$ & $0.0145\substack{+0.0013 \\ -0.0018}$ & $0.059\substack{+0.003 \\ -0.002}$ & $0.0647\substack{+0.0022 \\ -0.0024}$ \\ \hline
         $P({\rm Long~Bar})$ & $0.1323\substack{+0.0025 \\ -0.0025}$ & $0.201\substack{+0.007 \\ -0.007}$ & $0.206\substack{+0.007 \\ -0.010}$ \\ \hline
         $\sigma_r$ (kpc) & $4.92\substack{+0.04 \\ -0.03}$ & $5.66\substack{+0.05 \\ -0.04}$ & $6.07\substack{+0.05 \\ -0.08}$ \\ \hline
         $z_0$ (kpc) & $0.2238\substack{+0.0008 \\ -0.0027}$ & $1.203\substack{+0.023 \\ -0.013}$ & $1.222\substack{+0.018 \\ -0.004}$ \\ \hline
         $O_R$ (kpc) & $2.82\substack{+0.08 \\ -0.11}$ & $-$ & $-$ \\ \hline
         $O_N$ & $4.3\substack{+0.3 \\ -0.2}$ & $-$ & $-$ \\ \hline
         $\epsilon$ & $0.780\substack{+0.020 \\ -0.013}$ & $-$ & $-$ \\ \hline
         $a_b$ (kpc) & $0.205\substack{+0.009 \\ -0.010}$ & $0.509\substack{+0.019 \\ -0.010}$ & $0.555\substack{+0.022 \\ -0.021}$ \\ \hline
         $C_\perp$ & $1.84\substack{+0.02 \\ -0.05}$ & $1.831\substack{+0.013 \\ -0.015}$ & $1.821\substack{+0.017 \\ -0.015}$ \\ \hline
         $C_\parallel$ & $3.08\substack{+0.11 \\ -0.11}$ & $2.49\substack{+0.03 \\ -0.03}$ & $2.542\substack{+0.019 \\ -0.019}$ \\ \hline
         $x_b$ (kpc) & $0.557\substack{+0.008 \\ -0.008}$ & $0.685\substack{+0.005 \\ -0.005}$ & $0.673\substack{+0.007 \\ -0.011}$ \\ \hline
         $y_b$ (kpc) & $0.367\substack{+0.006 \\ -0.003}$ & $0.455\substack{+0.003 \\ -0.003}$ & $0.445\substack{+0.006 \\ -0.011}$ \\ \hline
         $z_b$ (kpc) & $0.2557\substack{+0.0014 \\ -0.0010}$ & $0.3328\substack{+0.0014 \\ -0.0015}$ & $0.3155\substack{+0.0026 \\ -0.0029}$ \\ \hline
         $R_{\rm end}$ (kpc) & $2.00\substack{+0.05 \\ -0.06}$ & $4.6\substack{+0.4 \\ -0.3}$ & $4.9\substack{+0.6 \\ -0.6}$ \\ \hline
         $\Delta \alpha$ ($\deg$) & $-$ & $-0.34\substack{+0.15 \\ -0.14}$ & $-0.07\substack{+0.13 \\ -0.13}$ \\ \hline
         $\Delta x_{\rm center}$ (pc) & $-$ & $0.8\substack{+1.0 \\ -1.1}$ & $-0.7\substack{+1.0 \\ -1.0}$ \\ \hline
         $\Delta y_{\rm center}$ (pc) & $-$ & $-1.8\substack{+0.9 \\ -0.9}$ & $-0.3\substack{+0.7 \\ -0.7}$ \\ \hline
         $\Delta z_{\rm center}$ (pc) & $-$ & $-0.9\substack{+0.6 \\ -0.5}$ & $2.7\substack{+0.4 \\ -0.5}$ \\ \hline
         $x_{\rm lb}$ (kpc) & $5\substack{+5 \\ -1}$ & $2.78\substack{+0.03 \\ -0.03}$ & $2.46\substack{+0.04 \\ -0.03}$ \\ \hline
         $y_{\rm lb}$ (kpc) & $1.43\substack{+0.16 \\ -0.07}$ & $2.78\substack{+0.04 \\ -0.04}$ & $2.49\substack{+0.05 \\ -0.04}$ \\ \hline
         $z_{\rm lb}$ (kpc) & $0.315\substack{+0.029 \\ -0.010}$ & $0.541\substack{+0.006 \\ -0.011}$ & $0.495\substack{+0.010 \\ -0.014}$ \\ \hline
         $C_{\perp,\rm lb}$ & $0.88\substack{+0.04 \\ -0.11}$ & $1.764\substack{+0.026 \\ -0.025}$ & $1.983\substack{+0.028 \\ -0.026}$ \\ \hline
         $R_{\rm out}$ (kpc) & $2.86\substack{+0.04 \\ -0.05}$ & $7.46\substack{+0.06 \\ -0.04}$ & $8.37\substack{+0.06 \\ -0.06}$ \\ \hline
         $R_{\rm in}$ (kpc) & $1.676\substack{+0.018 \\ -0.016}$ & $2.09\substack{+0.03 \\ -0.04}$ & $2.03\substack{+0.07 \\ -0.12}$ \\ \hline
    \end{tabular}
    \label{tab:MWa_params}
    \caption{ Fitted parameters for model MWa. We have used the median of the MCMC chains for the central value and also included 68\% confidence intervals.  }
    \end{center}
\end{table}

\begin{table}
\begin{center}
    \begin{tabular}{r||c|c|c}
         Parameter & No Kick & Kick At Beginning & Uniform Kick Rate \\ \hline \hline
         
         $P({\rm Disk})$ & $0.506\substack{+0.010 \\ -0.005}$ & $0.6144\substack{+0.0024 \\ -0.0026}$ & $0.36\substack{+0.03 \\ -0.04}$ \\ \hline
         $P({\rm Bar})$ & $0.322\substack{+0.005 \\ -0.003}$ & $0.244\substack{+0.007 \\ -0.003}$ & $0.317\substack{+0.006 \\ -0.003}$ \\ \hline
         $P({\rm Hernquist})$ & $0.0140\substack{+0.0012 \\ -0.0009}$ & $0.0470\substack{+0.0022 \\ -0.0021}$ & $0.0819\substack{+0.0017 \\ -0.0016}$ \\ \hline
         $P({\rm Long~Bar})$ & $0.154\substack{+0.012 \\ -0.010}$ & $0.094\substack{+0.004 \\ -0.006}$ & $0.24\substack{+0.04 \\ -0.04}$ \\ \hline
         $\sigma_r$ (kpc) & $5.64\substack{+0.06 \\ -0.07}$ & $5.36\substack{+0.03 \\ -0.03}$ & $6.9\substack{+0.7 \\ -0.3}$ \\ \hline
         $z_0$ (kpc) & $0.2249\substack{+0.0003 \\ -0.0003}$ & $1.116\substack{+0.004 \\ -0.014}$ & $1.30\substack{+0.05 \\ -0.06}$ \\ \hline
         $O_R$ (kpc) & $3.19\substack{+0.05 \\ -0.09}$ & $-$ & $-$ \\ \hline
         $O_N$ & $4.25\substack{+0.09 \\ -0.17}$ & $-$ & $-$ \\ \hline
         $\epsilon$ & $0.762\substack{+0.019 \\ -0.019}$ & $-$ & $-$ \\ \hline
         $a_b$ (kpc) & $0.202\substack{+0.004 \\ -0.005}$ & $0.585\substack{+0.023 \\ -0.025}$ & $0.614\substack{+0.014 \\ -0.011}$ \\ \hline
         $C_\perp$ & $2.04\substack{+0.08 \\ -0.10}$ & $2.02\substack{+0.03 \\ -0.03}$ & $1.873\substack{+0.016 \\ -0.017}$ \\ \hline
         $C_\parallel$ & $3.85\substack{+0.14 \\ -0.08}$ & $3.06\substack{+0.05 \\ -0.05}$ & $2.90\substack{+0.04 \\ -0.04}$ \\ \hline
         $x_b$ (kpc) & $0.548\substack{+0.029 \\ -0.009}$ & $0.587\substack{+0.026 \\ -0.009}$ & $0.710\substack{+0.010 \\ -0.006}$ \\ \hline
         $y_b$ (kpc) & $0.339\substack{+0.006 \\ -0.005}$ & $0.406\substack{+0.013 \\ -0.005}$ & $0.411\substack{+0.010 \\ -0.005}$ \\ \hline
         $z_b$ (kpc) & $0.243\substack{+0.004 \\ -0.002}$ & $0.306\substack{+0.007 \\ -0.003}$ & $0.303\substack{+0.004 \\ -0.002}$ \\ \hline
         $R_{\rm end}$ (kpc) & $2.53\substack{+0.10 \\ -0.15}$ & $3.37\substack{+0.10 \\ -0.08}$ & $5.2\substack{+0.6 \\ -0.4}$ \\ \hline
         $\Delta \alpha$ ($\deg$) & $-$ & $-0.04\substack{+0.13 \\ -0.13}$ & $0.30\substack{+0.11 \\ -0.12}$ \\ \hline
         $\Delta x_{\rm center}$ (pc) & $-$ & $4.4\substack{+1.2 \\ -1.2}$ & $2.5\substack{+1.2 \\ -1.3}$ \\ \hline
         $\Delta y_{\rm center}$ (pc) & $-$ & $1.8\substack{+0.9 \\ -0.8}$ & $-1.2\substack{+0.9 \\ -0.8}$ \\ \hline
         $\Delta z_{\rm center}$ (pc) & $-$ & $-0.6\substack{+0.7 \\ -0.8}$ & $1.4\substack{+0.5 \\ -0.5}$ \\ \hline
         $x_{\rm lb}$ (kpc) & $2.7\substack{+0.8 \\ -0.3}$ & $2.1\substack{+0.3 \\ -0.1}$ & $2.78\substack{+0.04 \\ -0.02}$ \\ \hline
         $y_{\rm lb}$ (kpc) & $1.14\substack{+0.17 \\ -0.05}$ & $0.93\substack{+0.08 \\ -0.03}$ & $2.65\substack{+0.07 \\ -0.04}$ \\ \hline
         $z_{\rm lb}$ (kpc) & $0.414\substack{+0.006 \\ -0.016}$ & $0.656\substack{+0.011 \\ -0.020}$ & $0.56\substack{+0.04 \\ -0.03}$ \\ \hline
         $C_{\perp,\rm lb}$ & $0.97\substack{+0.03 \\ -0.11}$ & $1.61\substack{+0.04 \\ -0.04}$ & $1.86\substack{+0.03 \\ -0.03}$ \\ \hline
         $R_{\rm out}$ (kpc) & $3.66\substack{+0.22 \\ -0.11}$ & $4.67\substack{+0.05 \\ -0.05}$ & $9.32\substack{+0.07 \\ -0.25}$ \\ \hline
         $R_{\rm in}$ (kpc) & $1.69\substack{+0.04 \\ -0.03}$ & $1.78\substack{+0.04 \\ -0.03}$ & $1.71\substack{+0.10 \\ -0.06}$ \\ \hline
    \end{tabular}
    \label{tab:MWb_params}
    \caption{ Fitted parameters for model MWb. }
    \end{center}
\end{table}

\begin{table}
\begin{center}
    \begin{tabular}{r||c|c|c}
         Parameter & No Kick & Kick At Beginning & Uniform Kick Rate \\ \hline \hline
         
         $P({\rm Disk})$ & $0.555\substack{+0.003 \\ -0.009}$ & $0.467\substack{+0.012 \\ -0.010}$ & $0.357\substack{+0.013 \\ -0.003}$ \\ \hline
         $P({\rm Bar})$ & $0.318\substack{+0.004 \\ -0.005}$ & $0.262\substack{+0.003 \\ -0.005}$ & $0.305\substack{+0.003 \\ -0.002}$ \\ \hline
         $P({\rm Hernquist})$ & $0.0099\substack{+0.0006 \\ -0.0005}$ & $0.0603\substack{+0.0021 \\ -0.0025}$ & $0.0619\substack{+0.0019 \\ -0.0020}$ \\ \hline
         $P({\rm Long~Bar})$ & $0.116\substack{+0.014 \\ -0.004}$ & $0.210\substack{+0.014 \\ -0.010}$ & $0.277\substack{+0.005 \\ -0.017}$ \\ \hline
         $\sigma_r$ (kpc) & $5.49\substack{+0.03 \\ -0.04}$ & $6.11\substack{+0.07 \\ -0.09}$ & $6.99\substack{+0.06 \\ -0.19}$ \\ \hline
         $z_0$ (kpc) & $0.2246\substack{+0.0007 \\ -0.0022}$ & $1.266\substack{+0.016 \\ -0.028}$ & $1.262\substack{+0.026 \\ -0.022}$ \\ \hline
         $O_R$ (kpc) & $2.5\substack{+0.4 \\ -0.16}$ & $-$ & $-$ \\ \hline
         $O_N$ & $4.9\substack{+2.6 \\ -1.1}$ & $-$ & $-$ \\ \hline
         $\epsilon$ & $0.82\substack{+0.12 \\ -0.12}$ & $-$ & $-$ \\ \hline
         $a_b$ (kpc) & $0.220\substack{+0.006 \\ -0.006}$ & $0.72\substack{+0.01 \\ -0.04}$ & $0.74\substack{+0.03 \\ -0.03}$ \\ \hline
         $C_\perp$ & $1.88\substack{+0.03 \\ -0.03}$ & $1.861\substack{+0.020 \\ -0.020}$ & $1.909\substack{+0.024 \\ -0.022}$ \\ \hline
         $C_\parallel$ & $3.37\substack{+0.05 \\ -0.05}$ & $2.69\substack{+0.03 \\ -0.04}$ & $2.69\substack{+0.03 \\ -0.03}$ \\ \hline
         $x_b$ (kpc) & $0.551\substack{+0.012 \\ -0.014}$ & $0.657\substack{+0.006 \\ -0.007}$ & $0.620\substack{+0.006 \\ -0.005}$ \\ \hline
         $y_b$ (kpc) & $0.342\substack{+0.007 \\ -0.008}$ & $0.434\substack{+0.005 \\ -0.007}$ & $0.378\substack{+0.005 \\ -0.003}$ \\ \hline
         $z_b$ (kpc) & $0.2395\substack{+0.0015 \\ -0.0018}$ & $0.3217\substack{+0.0023 \\ -0.0023}$ & $0.2845\substack{+0.0025 \\ -0.0018}$ \\ \hline
         $R_{\rm end}$ (kpc) & $2.03\substack{+0.07 \\ -0.10}$ & $4.81\substack{+0.29 \\ -0.25}$ & $5.5\substack{+2.1 \\ -0.9}$ \\ \hline
         $\Delta \alpha$ ($\deg$) & $-$ & $-0.79\substack{+0.18 \\ -0.18}$ & $0.32\substack{+0.13 \\ -0.13}$ \\ \hline
         $\Delta x_{\rm center}$ (pc) & $-$ & $1.4\substack{+1.4 \\ -1.3}$ & $1.0\substack{+1.2 \\ -1.1}$ \\ \hline
         $\Delta y_{\rm center}$ (pc) & $-$ & $-0.2\substack{+0.9 \\ -0.9}$ & $0.5\substack{+0.8 \\ -0.8}$ \\ \hline
         $\Delta z_{\rm center}$ (pc) & $-$ & $2.4\substack{+0.7 \\ -0.6}$ & $-0.5\substack{+0.5 \\ -0.5}$ \\ \hline
         $x_{\rm lb}$ (kpc) & $3\substack{+7 \\ -1}$ & $2.533\substack{+0.023 \\ -0.022}$ & $2.533\substack{+0.020 \\ -0.021}$ \\ \hline
         $y_{\rm lb}$ (kpc) & $1.5\substack{+0.4 \\ -0.2}$ & $2.584\substack{+0.027 \\ -0.025}$ & $2.417\substack{+0.021 \\ -0.022}$ \\ \hline
         $z_{\rm lb}$ (kpc) & $0.335\substack{+0.006 \\ -0.005}$ & $0.564\substack{+0.019 \\ -0.006}$ & $0.530\substack{+0.010 \\ -0.011}$ \\ \hline
         $C_{\perp,\rm lb}$ & $0.95\substack{+0.07 \\ -0.11}$ & $1.964\substack{+0.029 \\ -0.029}$ & $1.954\substack{+0.022 \\ -0.022}$ \\ \hline
         $R_{\rm out}$ (kpc) & $3.0\substack{+0.1 \\ -0.4}$ & $7.88\substack{+0.07 \\ -0.09}$ & $9.13\substack{+0.13 \\ -0.08}$ \\ \hline
         $R_{\rm in}$ (kpc) & $1.61\substack{+0.03 \\ -0.06}$ & $1.91\substack{+0.05 \\ -0.07}$ & $1.60\substack{+0.05 \\ -0.03}$ \\ \hline
    \end{tabular}
    \label{tab:MWc0.8_params}
    \caption{ Fitted parameters for model MWc0.8. }
    \end{center}
\end{table}

\begin{figure}
    \centering
    \includegraphics[width=0.99\linewidth]{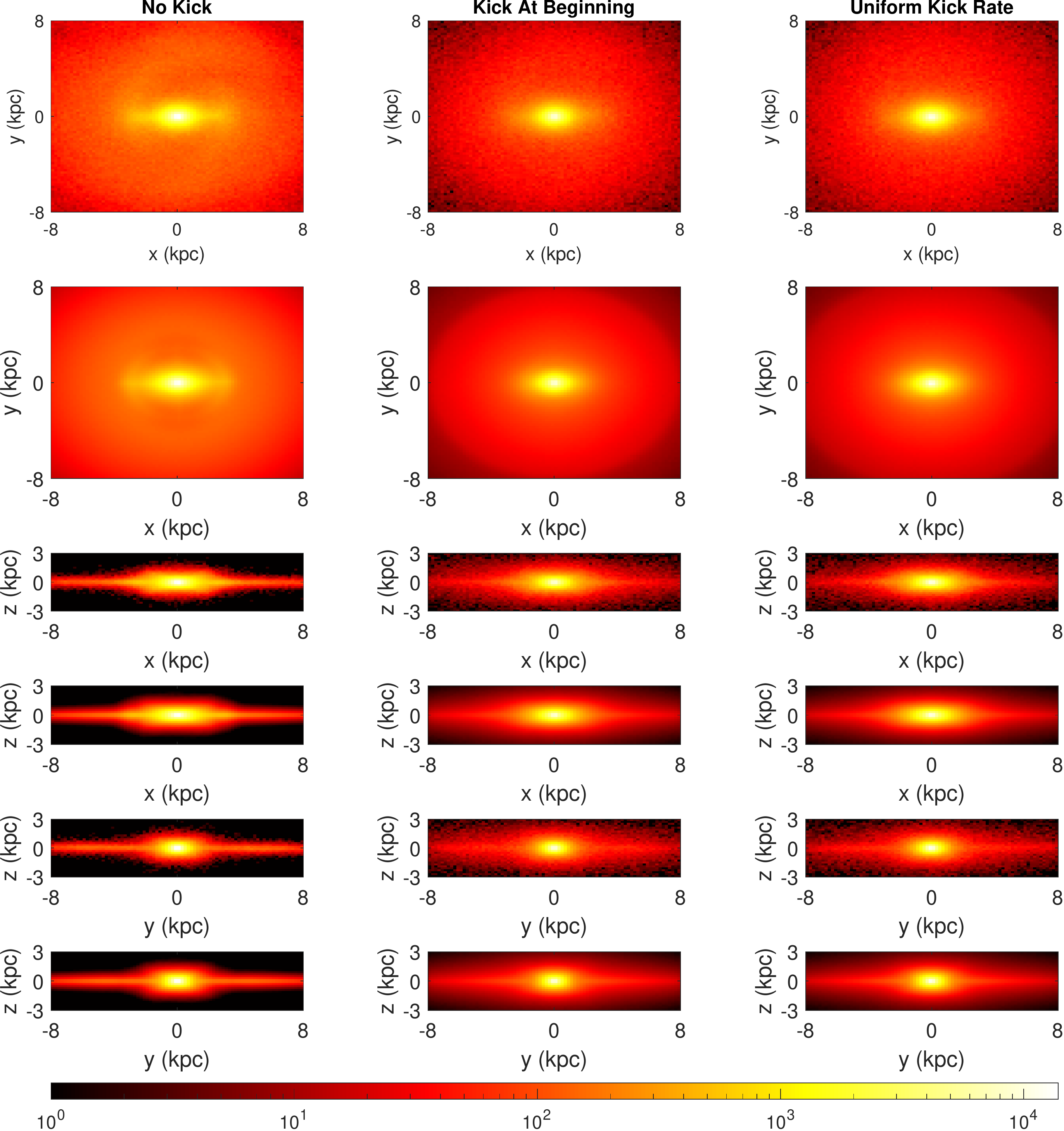}
    \caption{Final density map of particles with no kick, a kick at the beginning and a uniform kick rate generated from the MWa initial conditions. Every second row shows the fitted model. }
    \label{fig:MWa_density_data_fit}
\end{figure}

\begin{figure}
    \centering
    \includegraphics[width=0.99\linewidth]{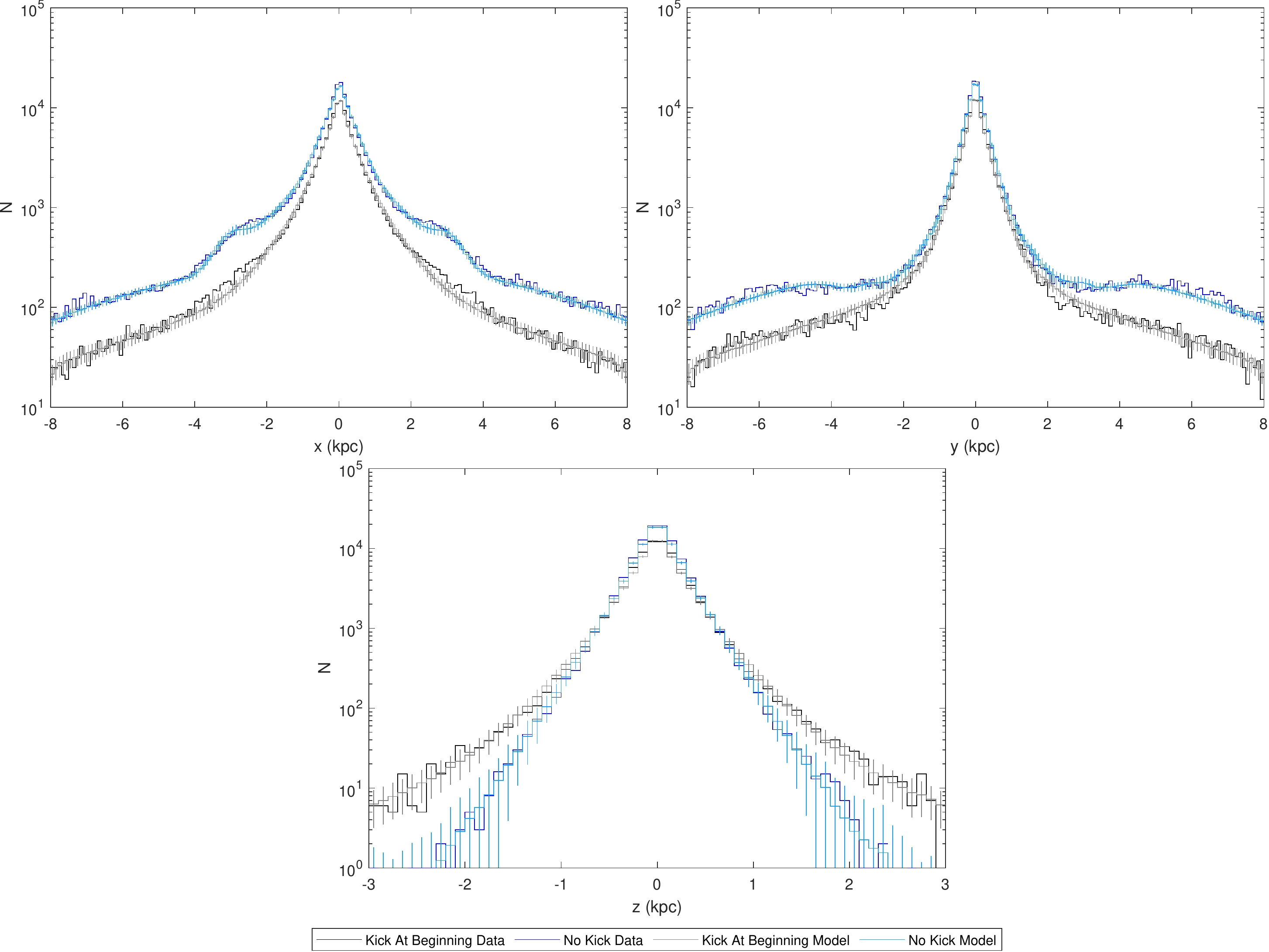}
    \caption{Profile along $x$, $y$ and $z$ axes with kicks occurring at the beginning for the case generated from the MWa initial conditions. We show both the final $N$-body simulation data and data simulated using the final fitted model. For the final fitted model we show the mean number of particles in each bin and the standard deviation. }
    \label{fig:MWa_1d_profiles_kick_at_beginning}
\end{figure}

\begin{figure}
    \centering
    \includegraphics[width=0.99\linewidth]{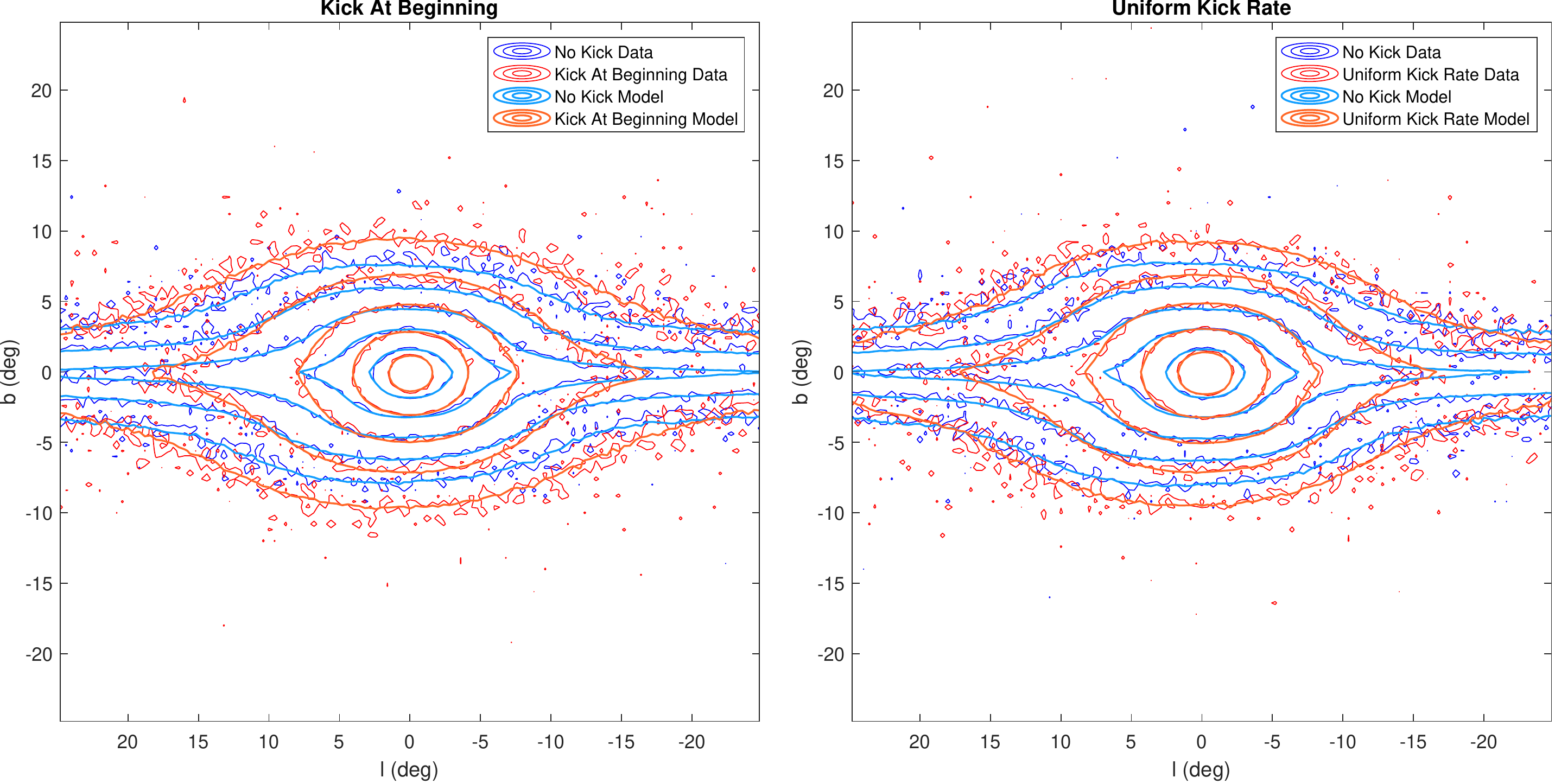}
    \caption{Final flux distribution in Galactic coordinates generated from the MWa initial conditions. The contours for each distribution are at $1$, $2$, $4$, $8$ and $16$ times the mean in this region. The Sun is placed at a distance of $7.9$ kpc, at an angle relative to the bar of $20^\circ$ and at a height of $15$ pc. }
    \label{fig:MWa_data_los}
\end{figure}

In Table~\ref{tab:alternative_bar} we show the change in $-2 \log(L)$ when replacing $\rho_{\rm bar} (R_s)$ with a range of different models from Freudenreich \cite{Freudenreich:1997bx} and Cao et al.~\cite{Cao:2013dwa}. As there is variation in the likelihood between chains, we also show the standard deviation in $-2 \log(L)$. Our choice of $\rho_{\rm bar} (R_s) \propto K_0(R_s)$ is clearly preferred over the others. The worst form, where $\rho_{\rm bar} (R_s) \propto \exp(R_s^{-n})$, was entirely removed with $P({\rm Bar}) = 0$ and the long bar component taking over the fit in the central region.

\begin{table}
\begin{center}
    \begin{tabular}{c||c|c}
          $\rho_{\rm bar} (R_s)$ & Mean $-2(\Delta \log(L))$ & Standard Deviation \\ \hline \hline
         
          $K_0(R_s)$ & $0$ & $739$ \\ \hline
          $\exp(-R_s)$ & $3581$ & $1606$ \\ \hline
          $\sech^2(R_s)$ & $12795$ & $1985$ \\ \hline
          $\exp(-0.5 R_s^2)$ & $19369$ & $391$ \\ \hline
          $(1 + R_s^n)^{-1}$ & $20403$ & $953$ \\ \hline
          $\exp(R_s^{-n})$ & $57114$ & $3152$ \\ \hline
    \end{tabular}
    \label{tab:alternative_bar}
    \caption{ Change in mean $-2 \log(L)$ using different bar models $\rho_{\rm bar} (R_s)$. The mean was taken over all the samples in the MCMC chains.}
    \end{center}
\end{table}

\begin{figure}
    \centering
    \includegraphics[width=0.8\linewidth]{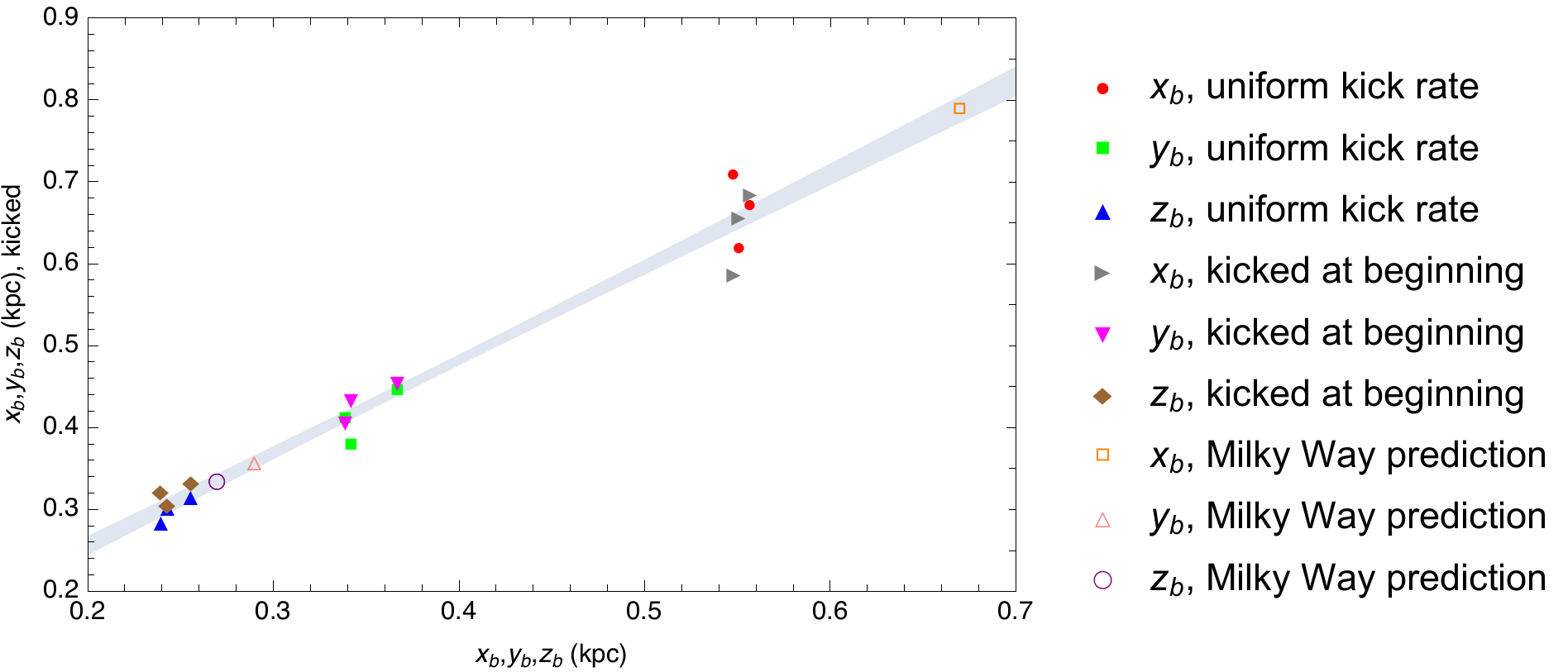}
    \includegraphics[width=0.8\linewidth]{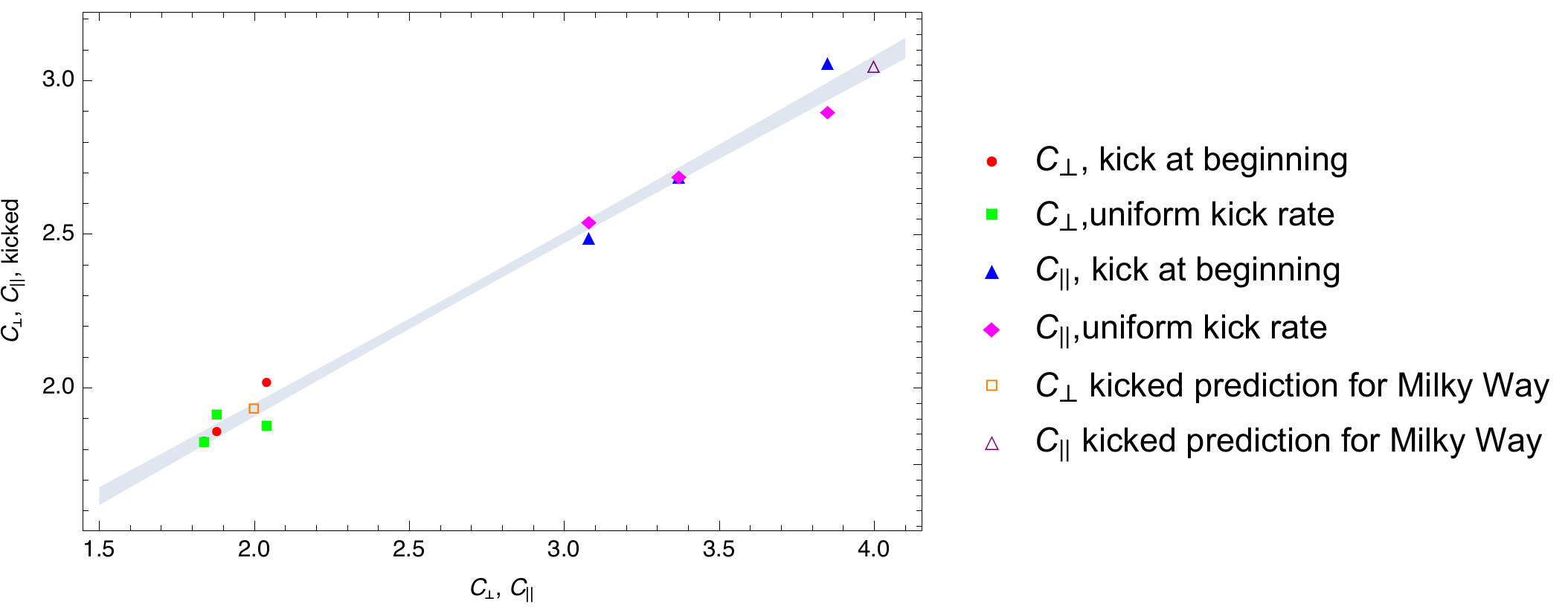}
    \caption{Simulation parameters with 68\% confidence interval bands for straight line model fits.
    The closed symbol values are obtained from values given in Tables~\ref{tab:MWa_params}, \ref{tab:MWb_params}, and \ref{tab:MWc0.8_params}.
    The predictions for the ref.~\cite{Cao:2013dwa} model of the Milky Way Galaxy are given as open symbols. 
    \label{fig:simulation_parameters}}
\end{figure}

\begin{figure}
    \centering
    \includegraphics[width=0.8\linewidth]{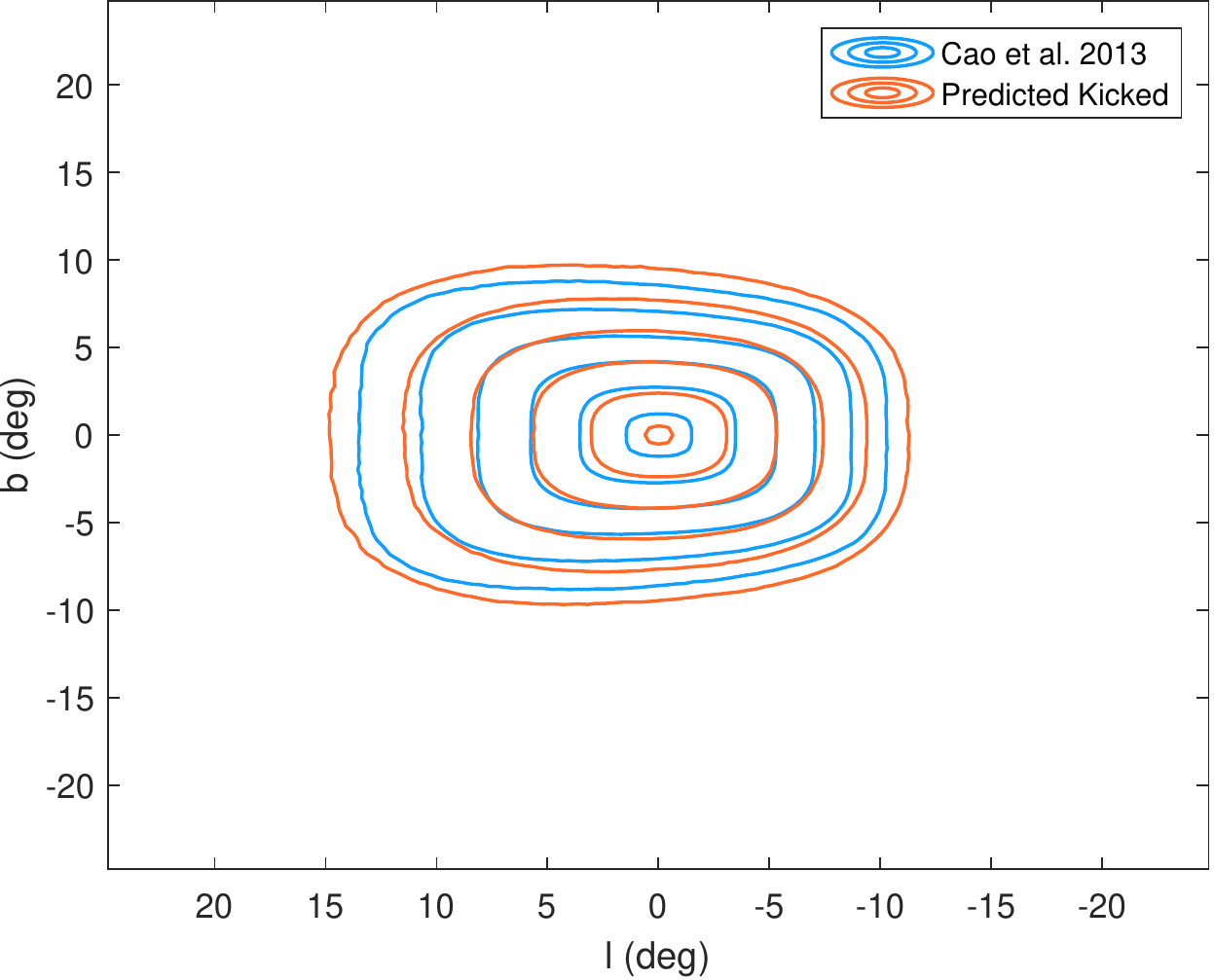}
    \caption{Line of sight contours for model of the Milky Way bar given by Cao et al.~\cite{Cao:2013dwa}. Both the model and its kicked version, obtained from the parameters in Table~\ref{tab:milky_way_prediction}, are shown. As in ref.~\cite{Cao:2013dwa} the Galactic Centre is taken to be 8.13 kpc away and the angle of the bulge to be 29$^\circ$. Contours are from 1 to 32 times the mean of the corresponding case given in steps of factors of 2.
    \label{fig:los_prediction}}
\end{figure}

In order to estimate the kick effects on the actual Milky Way Galactic bar we used a linear fit to our simulation results of the form
\begin{equation}
\theta_{{\rm kicked},i}=\alpha \theta_i +\beta
\label{eq:simulation_parameters}
\end{equation}
where $\theta_{{\rm kicked},i}$ are the bulge parameters, $x_b$, $y_b$, and $z_b$ for the kicked distribution  and 
$\theta_i$ are the corresponding parameters in the non-kicked case. The $\alpha$ and $\beta$ were found by performing a least squared fit for the values given in Tables~\ref{tab:MWa_params}, \ref{tab:MWb_params}, and \ref{tab:MWc0.8_params}.
We also did a similar fit for  $C_\perp$ and $C_\parallel$.
The results are shown in Table~\ref{tab:simulation_parameters} and Fig.~\ref{fig:simulation_parameters}. A prediction for the Milky Way bar parameters found in ref.~\cite{Cao:2013dwa} are shown in Table~\ref{tab:milky_way_prediction}. The predicted line of sight contours for the kicked and unkicked Milky Way bar are shown in Fig.~\ref{fig:los_prediction}.

\begin{table}
\begin{center}
    \begin{tabular}{r||c|c}
         Parameter & $\alpha_i$ & $\beta_i$ \\ \hline \hline
         
         $x_b, y_b, z_b$ & 1.13$\pm$0.05 &  0.03$\pm$0.02 \\ \hline
         $C_\perp,C_{\parallel}$& 0.56$\pm$0.02 &0.81$\pm$0.05\\
          \hline
    \end{tabular}
    \label{tab:simulation_parameters}
\caption{ Least square fit values with 68\% confidence intervals for Eq.~\ref{eq:simulation_parameters} fitted to the points shown in Figure~\ref{fig:simulation_parameters}. }
    \end{center}
\end{table}

\begin{table}
\begin{center}
    \begin{tabular}{r||c|c|c|c|c}
         Parameter & $x_b$ (kpc)   & $y_b$ (kpc)     & $z_b$ (kpc)   & $C_\perp$            & $C_{\parallel}$ \\ \hline \hline
         Not kicked & 0.67         &   0.29          &  0.27         &    2                 & 4        \\ \hline
           Kicked  & 0.79$\pm$0.02 & 0.36$\pm$0.01 & 0.35 $\pm$ 0.01 &     1.93$\pm$0.02 &  3.05$\pm$0.03       \\ 
          \hline
    \end{tabular}
    \label{tab:milky_way_prediction}
\caption{ Predictions with 68\% confidence intervals for the kicked spatial distribution for the Milky Way bar model found in ref.~\cite{Cao:2013dwa} using Eq.~\ref{eq:simulation_parameters} and the parameter values given in Table~\ref{tab:simulation_parameters}. }
    \end{center}
\end{table}

\section{Discussion and Conclusions}
\label{conclusion}

Our goal was to investigate the effect of neutron star birth kicks on the distribution of MSPs in the Galactic center.
We began by running $N$-body simulations with small populations of particles kicked with a range of scales in order to estimate the required Maxwellian kick to produce a peculiar velocity distribution similar to that of resolved gamma-ray MSPs. We then reran the simulations with a larger number of particles at the required kick velocity scale and used MCMC to fit the data with a model.

We used three initial condition models intended to approximate the Milky Way, these were the MWa, MWb and MWc0.8 initial condition models of Fujii et al.~\cite{Fujii2019}.
Our results were consistent with theirs as can be seen, for example, by comparing our Fig.~\ref{fig:rotation_curves}  to the top left hand panels of their Figs.~1, 2, and 3 which are also consistent with the observed circular rotation velocity at the Sun’s location.
In Cao et al.~\cite{Cao:2013dwa} the bar scale lengths for a modified Bessel function of the second kind model fitted to red clump giant data are $0.67$, $0.29$ and $0.27$ for the $x$, $y$ and $z$ axes respectively, with the parameters $C_\parallel$ and $C_\perp$ fixed at $4$ and $2$. Our fits to $N$-body models without kicks find $(x_b, y_b, z_b)$ of $(0.56, 0.37, 0.26)$ for MWa, $(0.55, 0.34, 0.24)$ for MWb and $(0.55, 0.34, 0.24)$ for MWc0.8. We also have $C_\parallel$ between $3$ and $4$, producing a boxy structure in $x$-$z$ and $y$-$z$, this is visible in Fig.~\ref{fig:MWa_density_data_fit} for the MWa case and Figs.~\ref{fig:MWb_density_data_fit} and \ref{fig:MWc0.8_density_data_fit} for the MWb and
MWc0.8 cases. The other shape parameter $C_\perp$ was relatively close to $2$ in all cases, resulting in a more elliptical shape in $x$-$y$.
We found it was necessary to extend the bar structure using the long bar component given in Eq.~\ref{eq:long_bar}. Without it the bar scale parameters would be larger, but the bar would not be long enough to explain  the structure seen for $2\mbox{ kpc} \lesssim \abs{x} \lesssim 4$ kpc in Fig.~\ref{fig:MWa_1d_profiles_kick_at_beginning}
in the MWa, kicked at the beginning case, and Figs.~\ref{fig:MWa_1d_profiles_uniform_kick_rate}, \ref{fig:MWb_1d_profiles_kick_at_beginning}, \ref{fig:MWb_1d_profiles_uniform_kick_rate}, \ref{fig:MWc0.8_1d_profiles_kick_at_beginning} and \ref{fig:MWc0.8_1d_profiles_uniform_kick_rate} in the uniform kick rate, MWb and MWc0.8 cases.
We find the disk scale height to be $0.22$~kpc for the MWa, MWb, and MWc0.8 unkicked cases. This is at the lower end of the range of $220$ to $450$ pc given in Bland-Hawthorn and Gerhard \cite{Bland-Hawthorn2016}. There is also a small spherically symmetric Hernquist component $\sim1\%$ of the particles in the region of interest for the unkicked cases.

\begin{figure}
    \centering
    \includegraphics[width=0.99\linewidth]{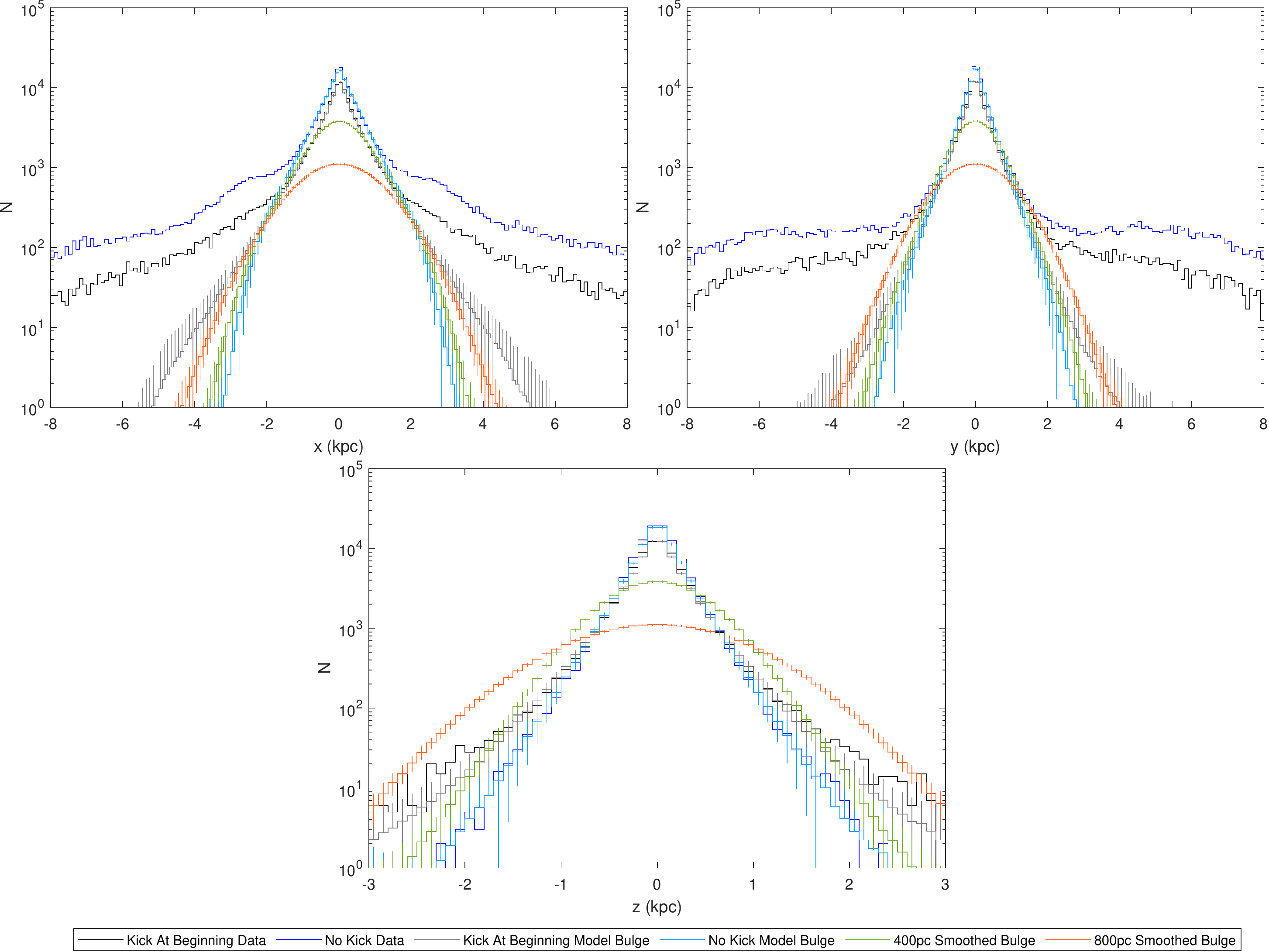}
    \caption{Final profile along $x$, $y$ and $z$ axes with kicks occurring at the beginning
    generated from the MWa initial conditions. Here we show the fitted bulge components, which consist of the bar plus Hernquist bulge,
    as well as the no kick bulge smoothed with $400$ pc and $800$ pc Gaussians. We show both $N$-body simulation data and data simulated using the fitted model. For the fitted model we show the mean number of particles in each bin and the standard deviation. }
    \label{fig:MWa_bulge_smoothed_1d_profiles_kick_at_beginning}
\end{figure}

\begin{figure}
    \centering
    \includegraphics[width=0.99\linewidth]{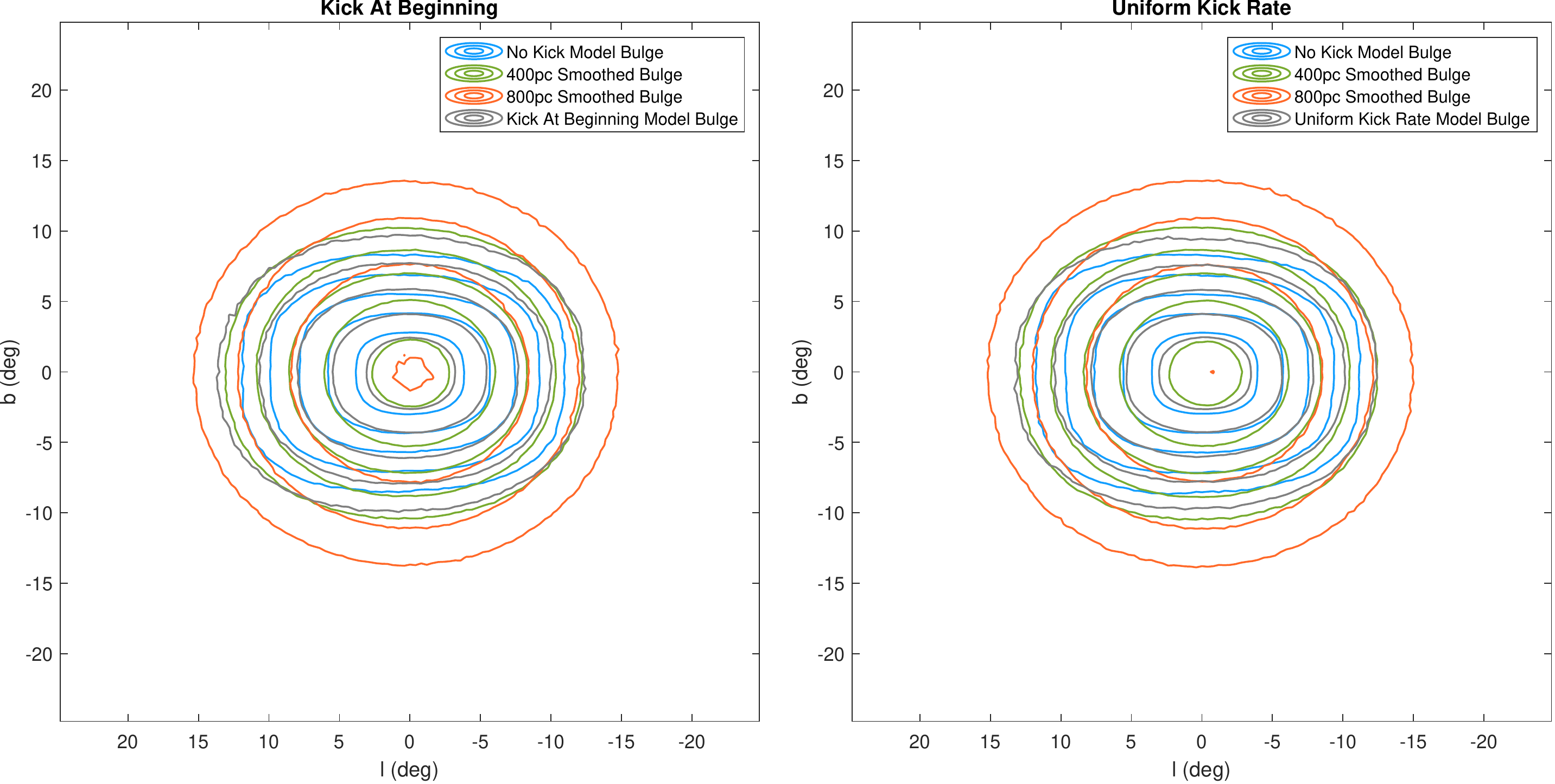}
    \caption{Final bulge, consisting of  bar plus Hernquist bulge, flux distribution in Galactic coordinates generated from the MWa initial conditions. We also show the no kick bulge smoothed with $400$ pc and $800$ pc Gaussians. The contours for each distribution are at $1$, $2$, $4$, $8$ and $16$ times the mean in this region. The Sun is placed at a distance of $7.9$ kpc, at an angle relative to the bar of $20^\circ$ and at a height of $15$ pc. }
    \label{fig:MWa_bulge_smoothed_los}
\end{figure}

\begin{figure}
    \centering
    \includegraphics[width=0.99\linewidth]{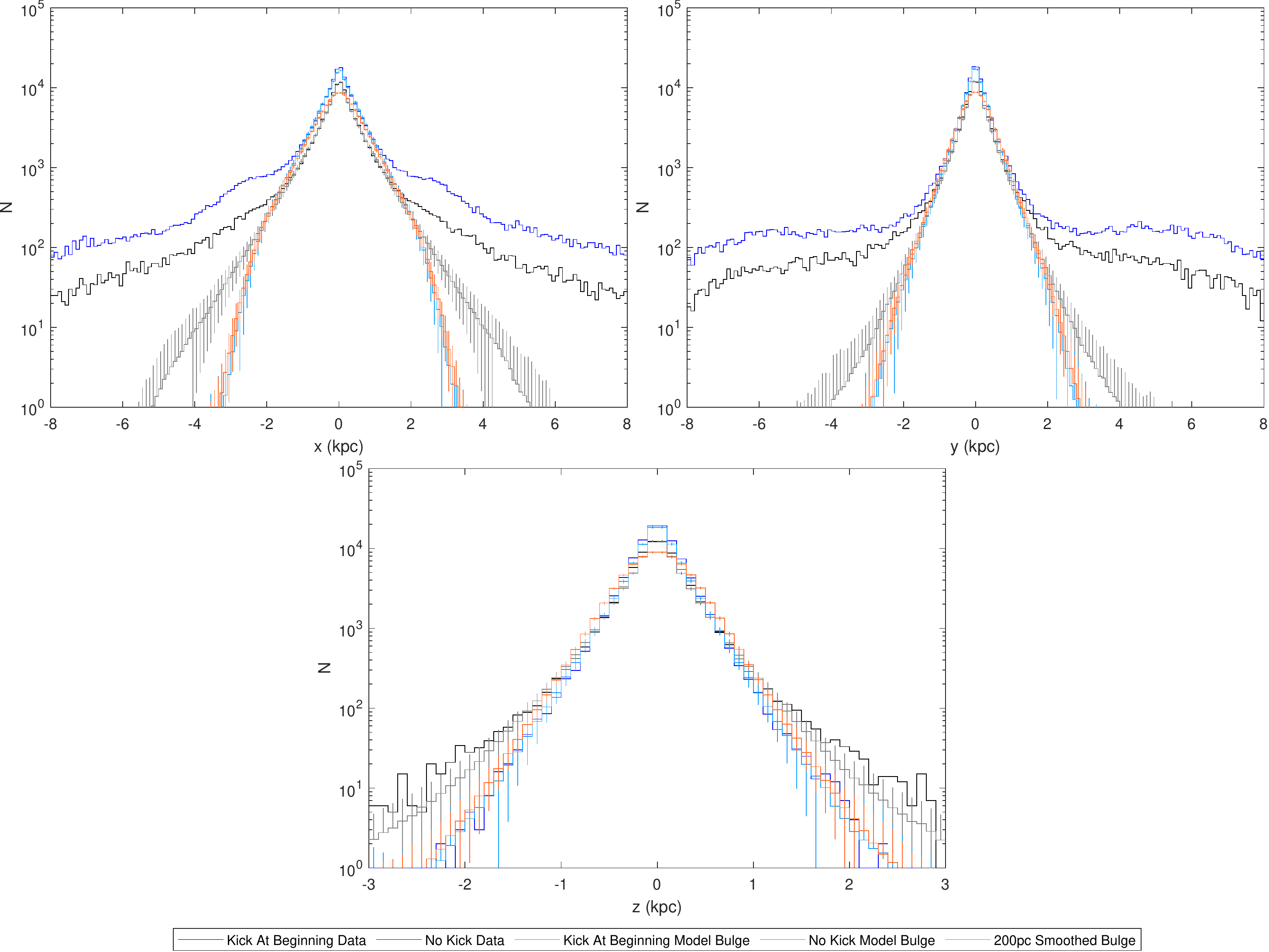}
    \caption{Final profile along $x$, $y$ and $z$ axes with kicks occurring at the beginning generated from the MWa initial conditions. Here we show the fitted bulge components as well as the no kick bulge smoothed with a $200$ pc Gaussian. We show both $N$-body simulation data and data simulated using the fitted model. For the fitted model we show the mean number of particles in each bin and the standard deviation. }
    \label{fig:MWa_bulge_smoothed_200pc_1d_profiles_kick_at_beginning}
\end{figure}

\begin{figure}
    \centering
    \includegraphics[width=0.99\linewidth]{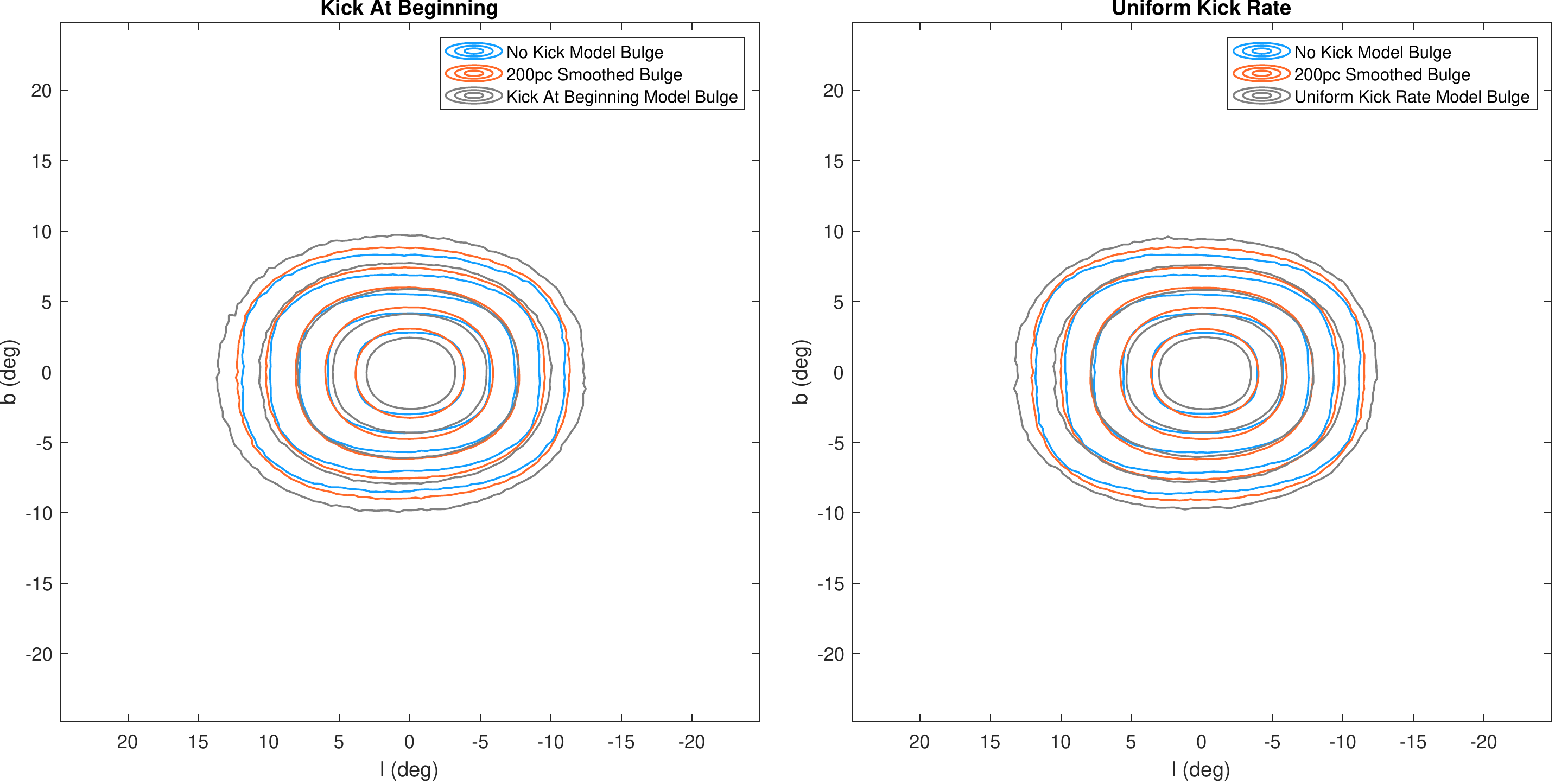}
    \caption{Final bulge flux distribution in Galactic coordinates generated from the MWa initial conditions. We also show the no kick bulge smoothed with a $200$ pc Gaussian. The contours for each distribution are at $1$, $2$, $4$, $8$ and $16$ times the mean in this region. The Sun is placed at a distance of $7.9$ kpc, at an angle relative to the bar of $20^\circ$ and at a height of $15$ pc. }
    \label{fig:MWa_bulge_smoothed_200pc_los}
\end{figure}

\begin{figure}
    \centering
    \includegraphics[width=0.99\linewidth]{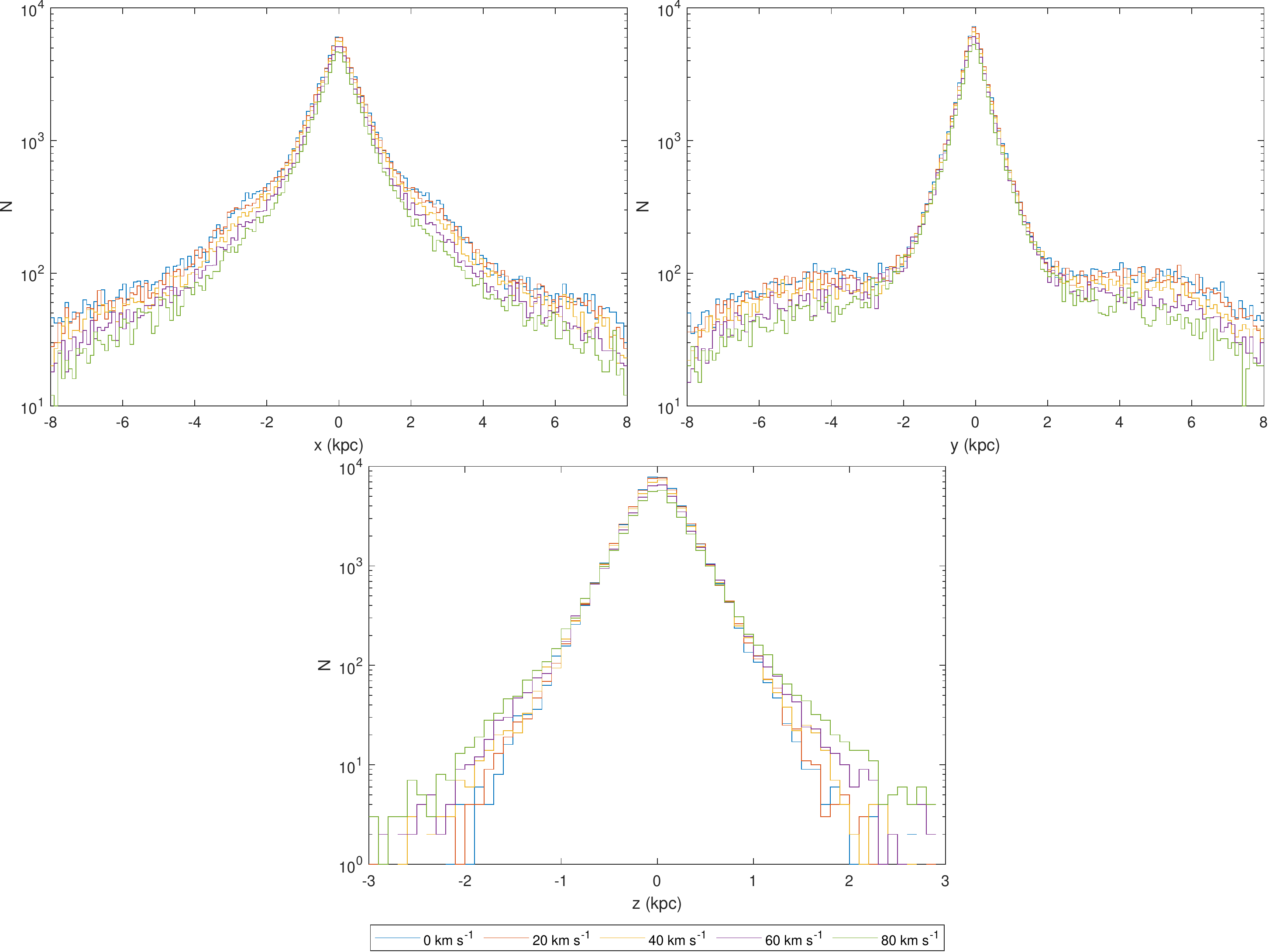}
    \caption{Final profile along $x$, $y$ and $z$ axes with kicks occurring at the beginning for kicks between $0$ km s$^{-1}$ and $80$ km s$^{-1}$ in $20$ km s$^{-1}$ increments. Generated from the MWa initial conditions. }
    \label{fig:MWa_less_80km_per_s_kick_1d_profiles_kick_at_beginning}
\end{figure}

Eckner et al.~\cite{Eckner2018} argued using the virial theorem that kicks $\langle v^2 \rangle \lesssim (70 {\rm km~s^{-1}})^2$ would lead to a ``smoothing" of the distribution of $700$--$900$ pc.
We show the effect of a $400$ pc and an $800$ pc Gaussian smoothing on the fitted bulge (bar plus Hernquist bulge) distribution in Figs.~\ref{fig:MWa_bulge_smoothed_1d_profiles_kick_at_beginning} and \ref{fig:MWa_bulge_smoothed_los} for the kicked at the beginning case.
The uniform kick rate case is shown in Fig.~\ref{fig:MWa_bulge_smoothed_1d_profiles_uniform_kick_rate}.
Those figures also show the bulge component of the models with and without kicks for comparison. It is clear that a Gaussian smoothing kernel will remove the peak that survives in the $N$-body simulations of kicked distributions and, particularly for the $800$ pc case, will produce an apparently spherically symmetric bulge. 
From the peculiar velocity data we inferred kicks that are larger than assumed by Eckner et al.~\cite{Eckner2018} with $\sigma_k$ at around $80$--$100$ km s$^{-1}$ ($\langle v^2 \rangle = 3 \sigma_k^2$ for a Maxwell distribution where the angular brackets signify the mean value) so the smoothing effect of the Gaussian would be even more severe.
We show an even smaller Gaussian smoothing of $200$ pc in Figs.~\ref{fig:MWa_bulge_smoothed_200pc_1d_profiles_kick_at_beginning} and \ref{fig:MWa_bulge_smoothed_200pc_los}
for the kicked at beginning case and in Fig.~\ref{fig:MWa_bulge_smoothed_200pc_1d_profiles_uniform_kick_rate} for the uniform kick rate case. We also show in Fig.~\ref{fig:MWa_less_80km_per_s_kick_1d_profiles_kick_at_beginning}  the profile for particles with smaller kick scales between $0$ km s$^{-1}$ and $80$ km s$^{-1}$ for the kicked at the beginning case. The corresponding uniform kick rate case is shown in Fig.~\ref{fig:MWa_less_80km_per_s_kick_1d_profiles_uniform_kick_rate}. In these two figures, each kick scale has only $4\times10^5$ particles; therefore, to reduce noise, the bins in each of the other two dimensions are twice as big as in previous single dimensional plots, and particles within $0.5$ kpc (previously $0.25$ kpc) of the axis are included. The profiles along the $z$ axis in particular show that there is a reduction in the slope as the kick velocities increase, along the other two axes the general increase in scaleheight is seen as a reduction in density.
These results demonstrates that Gaussian smoothing is not a good way of modelling a kicked version of a boxy bulge/bar template.

In every case, the bar fitted to the kicked data is both broader, with larger scale parameters $x_b$, $y_b$ and $z_b$, and less boxy, with smaller $C_\parallel$. For example, for the MWa  initial conditions $(x_b, y_b, z_b)$ increases from $(0.56, 0.37, 0.26)$ to $(0.69, 0.46, 0.33)$ and $(0.67, 0.45, 0.32)$ for the kick at beginning case and the uniform kick rate case respectively, while $C_\parallel$ declines from $3.08$ to $2.49$ and $2.54$. The spherically symmetric Hernquist bulge increases from $\sim1\%$ of the particles to $6\%$ for MWa. Like the bar, it becomes broader with $a_b$ increasing from around $0.2$ kpc to $0.51$ kpc and $0.56$ kpc. For the other two models similar changes occur, $P({\rm Long~Bar})$ and $a_b$ both increase significantly. In MSP model A1 of P20, the disk parameters were $\sigma_r = 4.5\substack{+0.5 \\ -0.4}$ kpc and $z_0 = 0.71\substack{+0.11 \\ -0.09}$. In the current article, after being kicked, the disk scale heights $z_0$ of all models increase from $0.22$ kpc to $\gtrsim{1}$ kpc, while $\sigma_r$ is in the range $5$--$7$~kpc. However, we find that in all kicked cases, except for MWb with kicks occurring at the beginning, the long bar behaves like a relatively thin disk component. We have $x_{\rm lb} \approx y_{\rm lb}$, $C_{\perp,\rm lb} \sim 2$ and $R_{\rm out} \gtrsim{7}$, resulting in a density $\sim \exp(-R/R_0)$ in $R$ for scalelength $R_0$. These scalelengths would then be between approximately $2.4$ kpc and $2.8$ kpc. For comparison, in Bland-Hawthorn and Gerhard \cite{Bland-Hawthorn2016} the Milky Way disk scalelength is reported as $2.6 \pm 0.5$ kpc. The exponential scaleheights of these ``long bars" range between about $0.5$--$0.6$ kpc. In Fig.~\ref{fig:MWa_1d_profiles_kick_at_beginning}, for the MWa kicked at the beginning case and in Figs.~\ref{fig:MWa_1d_profiles_uniform_kick_rate}, \ref{fig:MWb_1d_profiles_uniform_kick_rate}, \ref{fig:MWc0.8_1d_profiles_kick_at_beginning} and  \ref{fig:MWc0.8_1d_profiles_uniform_kick_rate} for the uniform kick rate and MWb, and MWc0.8 case,
there may be, to varying degrees, an excess of kicked particles over the model in the region of $2\mbox{ kpc} \lesssim \abs{x} \lesssim 4$ kpc.
 \newtext{This can be seen as a correlated run of roughly two  standard deviation difference between the prediction and the data in this region.}

\newtext{Although, a more complicated parametric model could remedy these fit defects, } our main aim in this article was to estimate the effect of the MSP kicks on their distribution in the Milky Way Galactic bar.
\newtext{As can be seen for the MWa case from Fig.~\ref{fig:MWa_data_los}, the line of sight integral of the models are a good match to the simulations. In particular the noisy simulation contours scatter in an unbiased way around the smooth model contours. Similar results can be seen for the MWb and MWc0.8 cases in Figs.~\ref{fig:MWb_data_los} and \ref{fig:MWc0.8_data_los} respectively.}

\newtext{We used the same bar parametric model as Cao et al.~\cite{Cao:2013dwa} who fit the best fit parameters to the  red clump 
luminosity density distribution of the Galactic bar measured by the Optical Gravitational Lensing Experiment (OGLE) III
survey. 
Comparing to their fit, our final simulation fits of the unkicked particles had somewhat different bulge parameters.} But, there appears to be a linear relationship between the unkicked scale parameters $x_b$, $y_b$, and $z_b$ and their kicked counterparts. Similarly, there appears to be a linear relationship between $C_\perp$, $C_{\parallel}$, and their kicked counterparts.
Therefore, we were able to estimate the Milky Way Galactic bar kicked parameters as shown in  Fig.~\ref{fig:simulation_parameters} and Table~\ref{tab:milky_way_prediction}. \newtext{The residuals in this least squares fit are partially due to the systematic error of our model misfit and also the Cao et al.~\cite{Cao:2013dwa} model misfit to the 
red clump data.}
As can be seen, there is more scatter in the $x_b$ parameter. This is not unexpected due to the already mentioned degeneracy with the long bar.
Also, as can be seen, estimating the Milky Way bar kicked $x_b$ parameter did involve a reasonable amount of extrapolation and so future simulations which have a larger $x_b$ will be needed to check it. 
A made-to-measure \cite{SyerTremaine1996,deLorenzi2007} approach may be needed. This would also be advantageous as it could take into account the X/peanut shaped morphology of the bar \cite{Nataf:2010,Mc10,Wegg2015} as done in ref.~\cite{PortailWeggGerhard2015}. There is some preliminary evidence that the X-shape may improve the fit to the Fermi-LAT gamma-ray data \cite{Coleman19}.

In conclusion, we used $N$-body simulations to explore the effect of a Maxwell distributed kick on the distribution of MSPs in the Galactic center. We find that while a $700$--$900$ pc Gaussian smoothing of the stellar mass would be too aggressive, the bulge distribution of the kicked particles is slightly broader and less boxy. From these results, we expect the GCE to deviate by a small amount from the stellar mass spatial distribution in the Galactic center. Also, as can seen from Table~\ref{tab:milky_way_prediction}, we would not expect the GCE to appear spherically symmetric due to the MSP kicks as that would require $x_b=y_b=z_b$ and $C_\perp=C_\parallel=2$ which are far from our inferred points relative to their error bars.

The amount  of spatial smoothing of the bulge MSPs will depend on the proportion of MSPs in the bulge that are made from the recycling channel and the proportion that are made from the 
 accretion induced collapse channel. Motivated by similarities between the bulge and disk population seen by P20 we have assumed this mixture has the same proportions as the disk MSPs. 
 If the GCE is due to bulge MSPs, its morphology could be used to
 check our smoothing prediction by comparing if there are any deviations between the GCE morphology and the stellar spatial distribution.
 A complication to this approach would be the possibility of some smearing of the GCE due to cosmic ray electron diffusion \cite{Song2019,Macias_2021}.
 An additional complication is that 
 if the MSP is spun up by a captured star then the MSP spatial distribution would be proportional to the stellar density squared \cite{Eckner2018, Macias19}.
 We have been assuming that the MSPs formed in a  binary system and so have a spatial distribution proportional to the stellar density. 
 Eventually, once the bulge MSPs are resolved \cite{Ploeg2020,Calore2016}, comparing their spatial distribution to the stellar distribution should provide independent information to more robustly estimate the natal kick distribution.

\section*{Acknowledgments}
We thank Roland Crocker and Oscar Macias for helpful comments. HP was supported by a University of Canterbury Doctoral Scholarship. 

\bibliographystyle{JHEP}
\bibliography{references}
\setcounter{figure}{0} 
\FloatBarrier
\appendix
\section{Uniform kick rate,  MWb, and MWc0.8 figures}
\counterwithin{figure}{section}
As the uniform kick rate,  MWb, and MWc0.8 gave similar results to the MWa kicked at beginning case we have moved many of their figures to this appendix.

\begin{figure}
    \centering
    \includegraphics[width=0.99\linewidth]{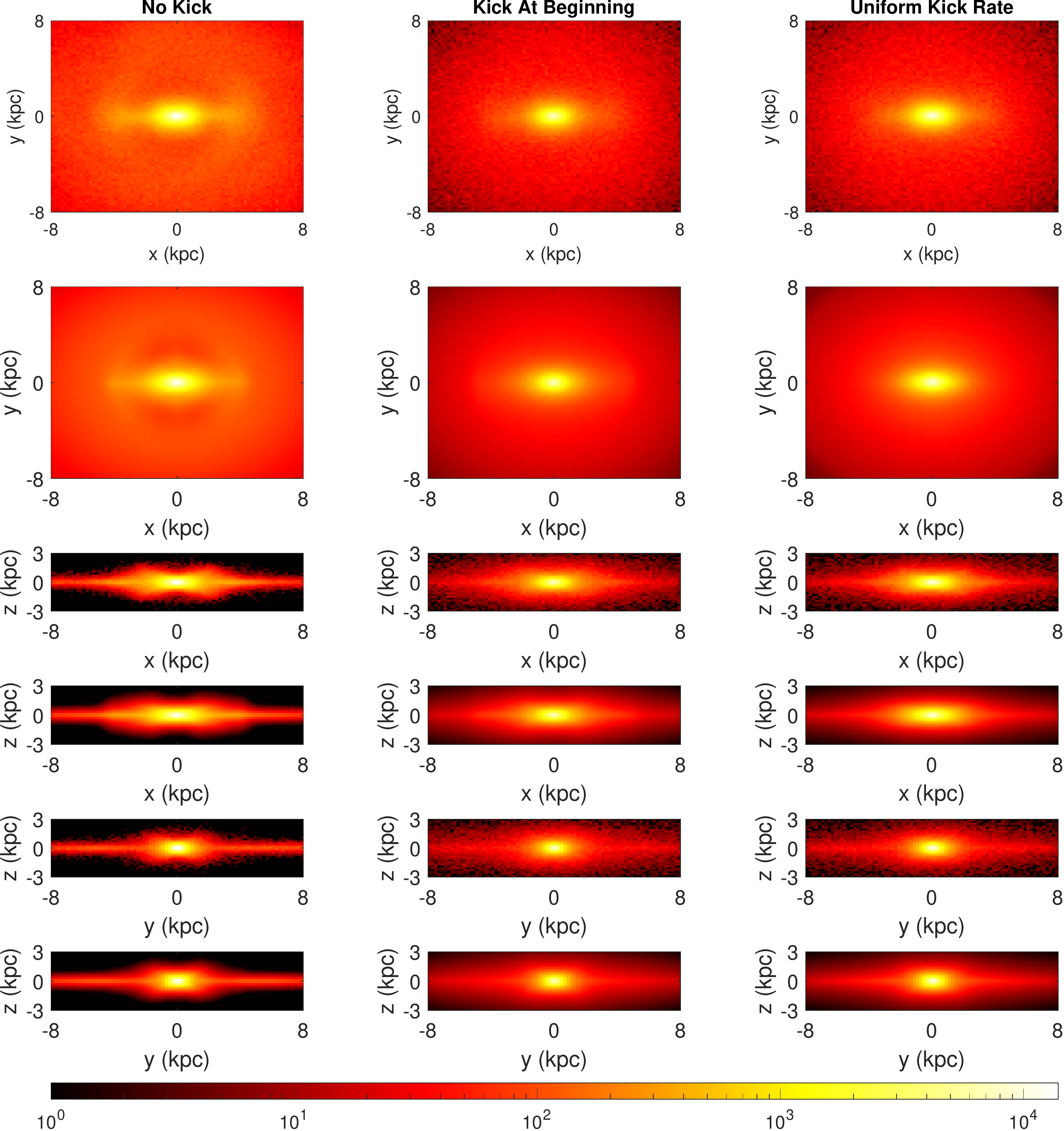}
    \caption{Final density map of particles with no kick, a kick at the beginning and a uniform kick rate generated from the MWb initial conditions. Every second row shows the fitted model. }
    \label{fig:MWb_density_data_fit}
\end{figure}

\begin{figure}
    \centering
    \includegraphics[width=0.99\linewidth]{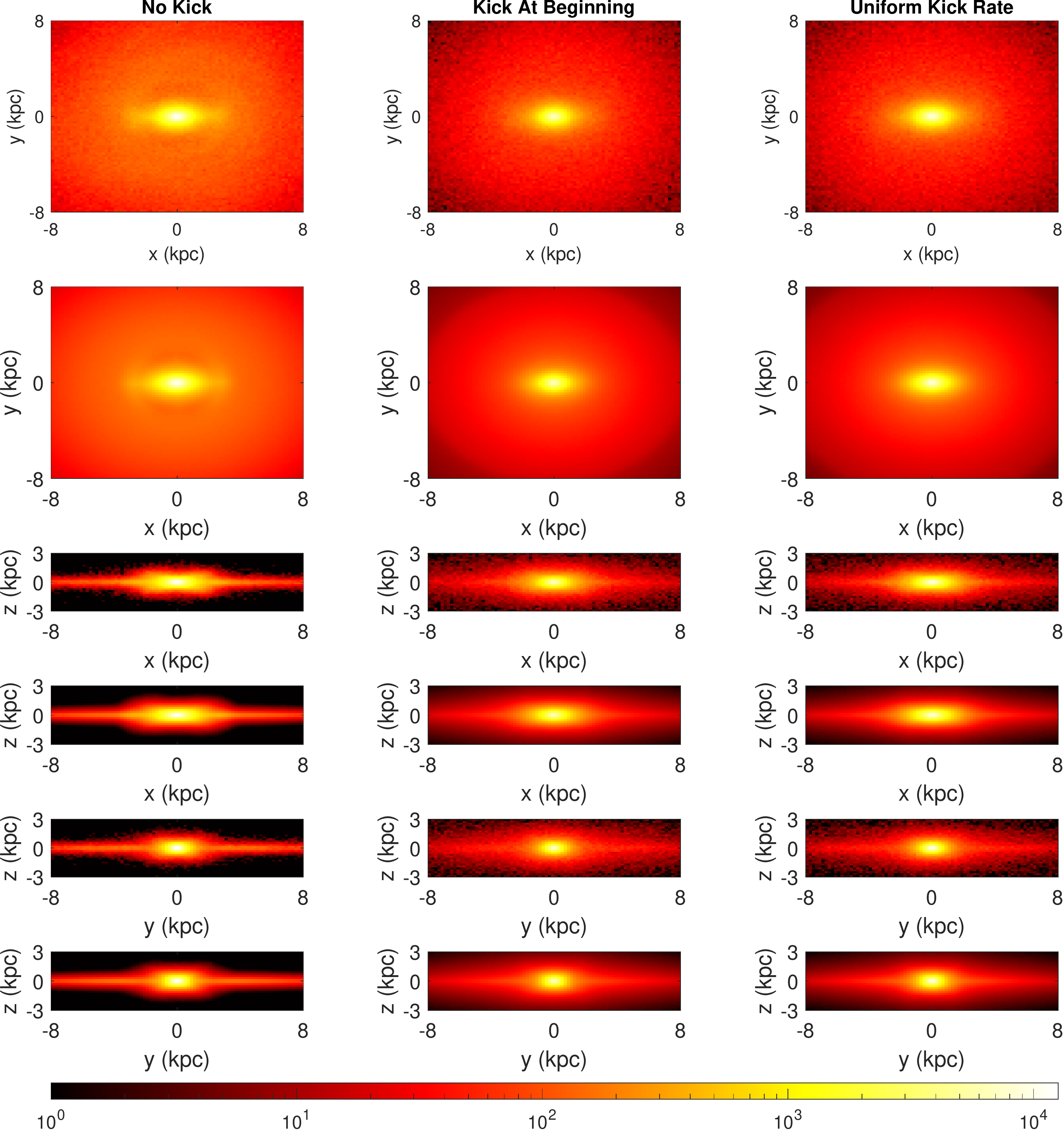}
    \caption{Final density map of particles with no kick, a kick at the beginning and a uniform kick rate generated from the  MWc0.8 initial conditions. Every second row shows the fitted model. }
    \label{fig:MWc0.8_density_data_fit}
\end{figure}

\begin{figure}
    \centering
    \includegraphics[width=0.99\linewidth]{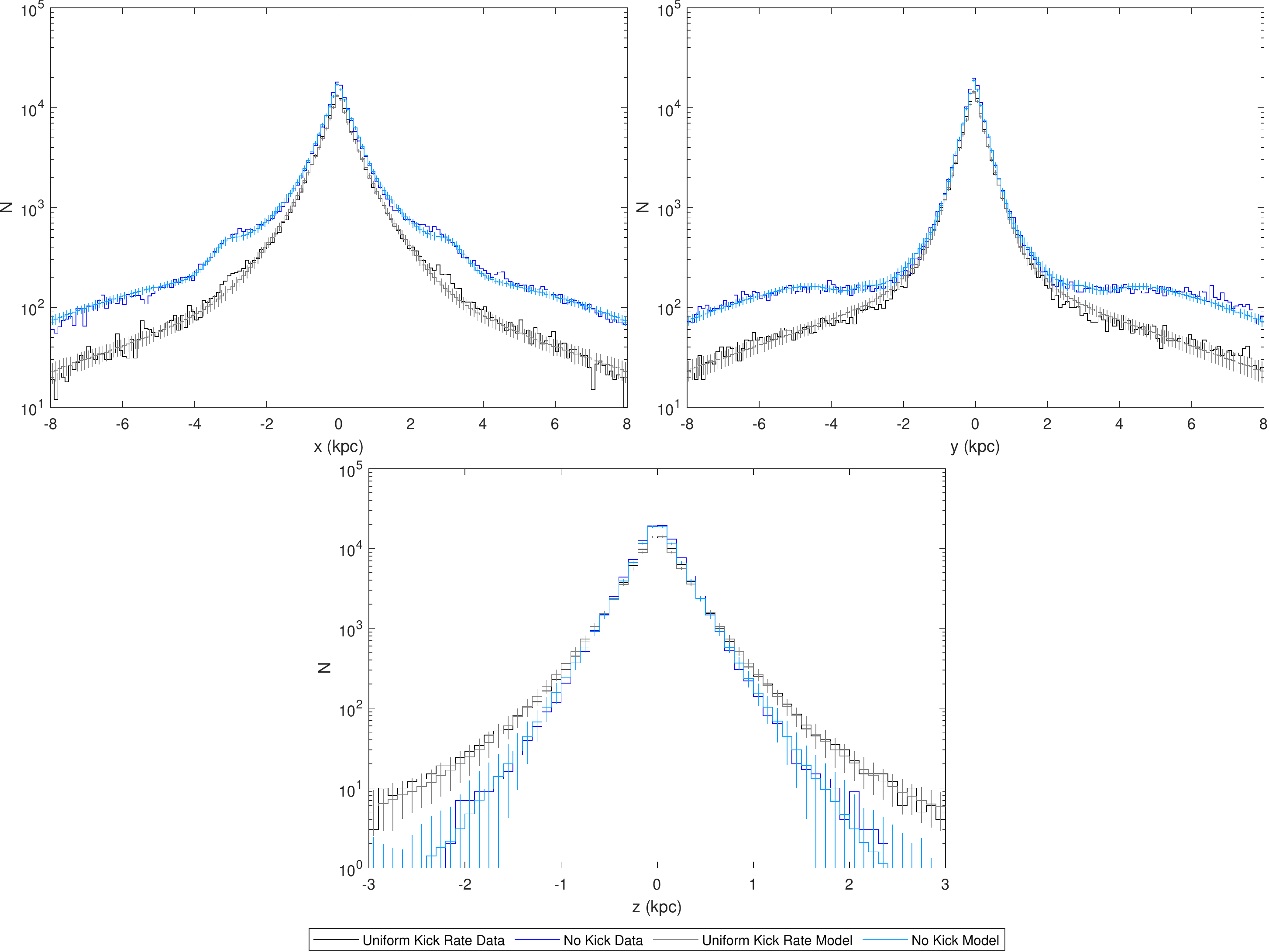}
    \caption{Final profile along $x$, $y$ and $z$ axes with a uniform kick rate generated from the MWa initial conditions. We show both $N$-body simulation data and data simulated using the fitted model. For the fitted model we show the mean number of particles in each bin and the standard deviation. }
    \label{fig:MWa_1d_profiles_uniform_kick_rate}
\end{figure}

\begin{figure}
    \centering
    \includegraphics[width=0.99\linewidth]{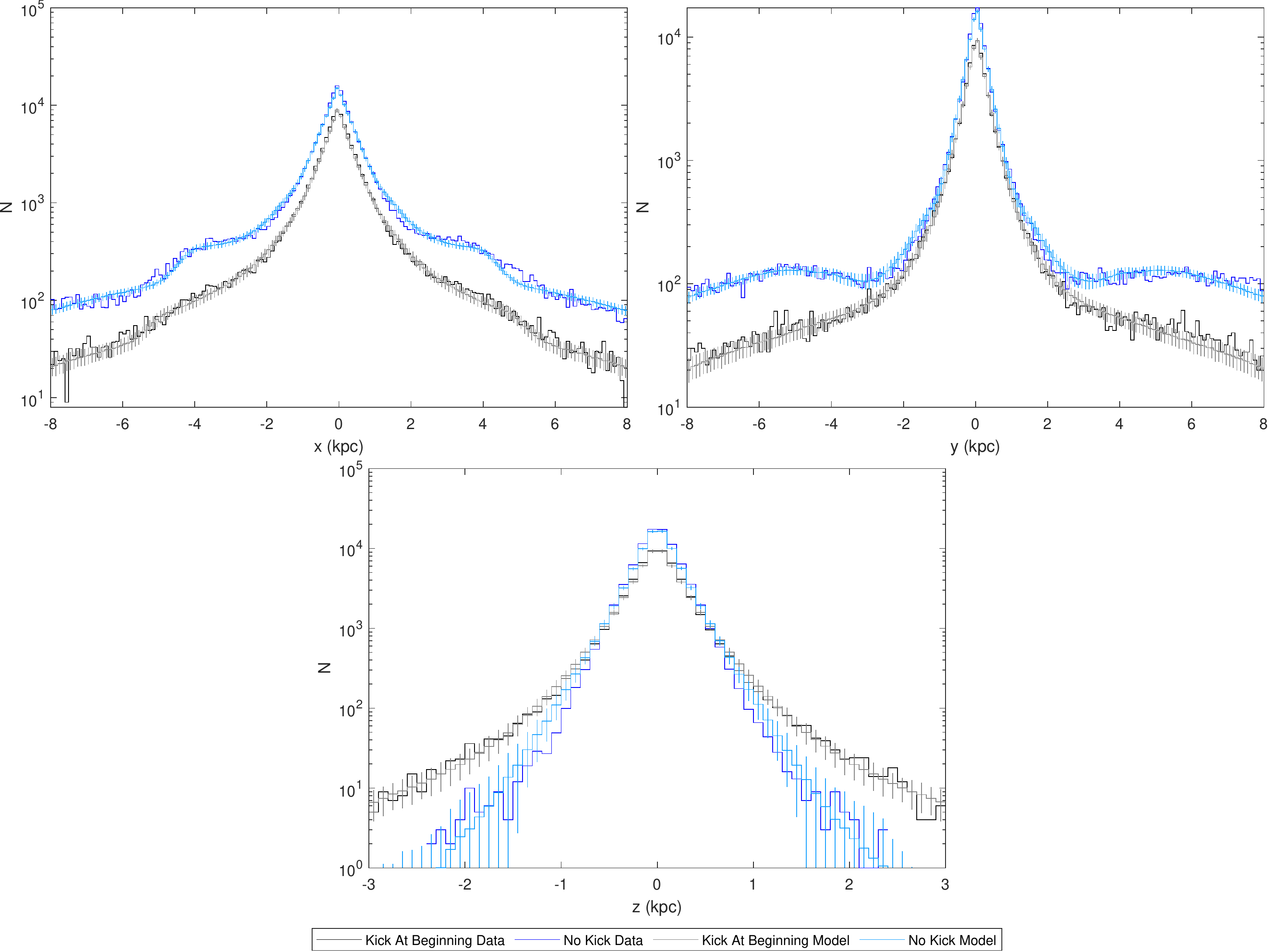}
    \caption{Final profile along $x$, $y$ and $z$ axes with kicks occurring at the beginning generated from the MWb initial conditions. We show both $N$-body simulation data and data simulated using the fitted model. For the fitted model we show the mean number of particles in each bin and the standard deviation. }
    \label{fig:MWb_1d_profiles_kick_at_beginning}
\end{figure}

\begin{figure}
    \centering
    \includegraphics[width=0.99\linewidth]{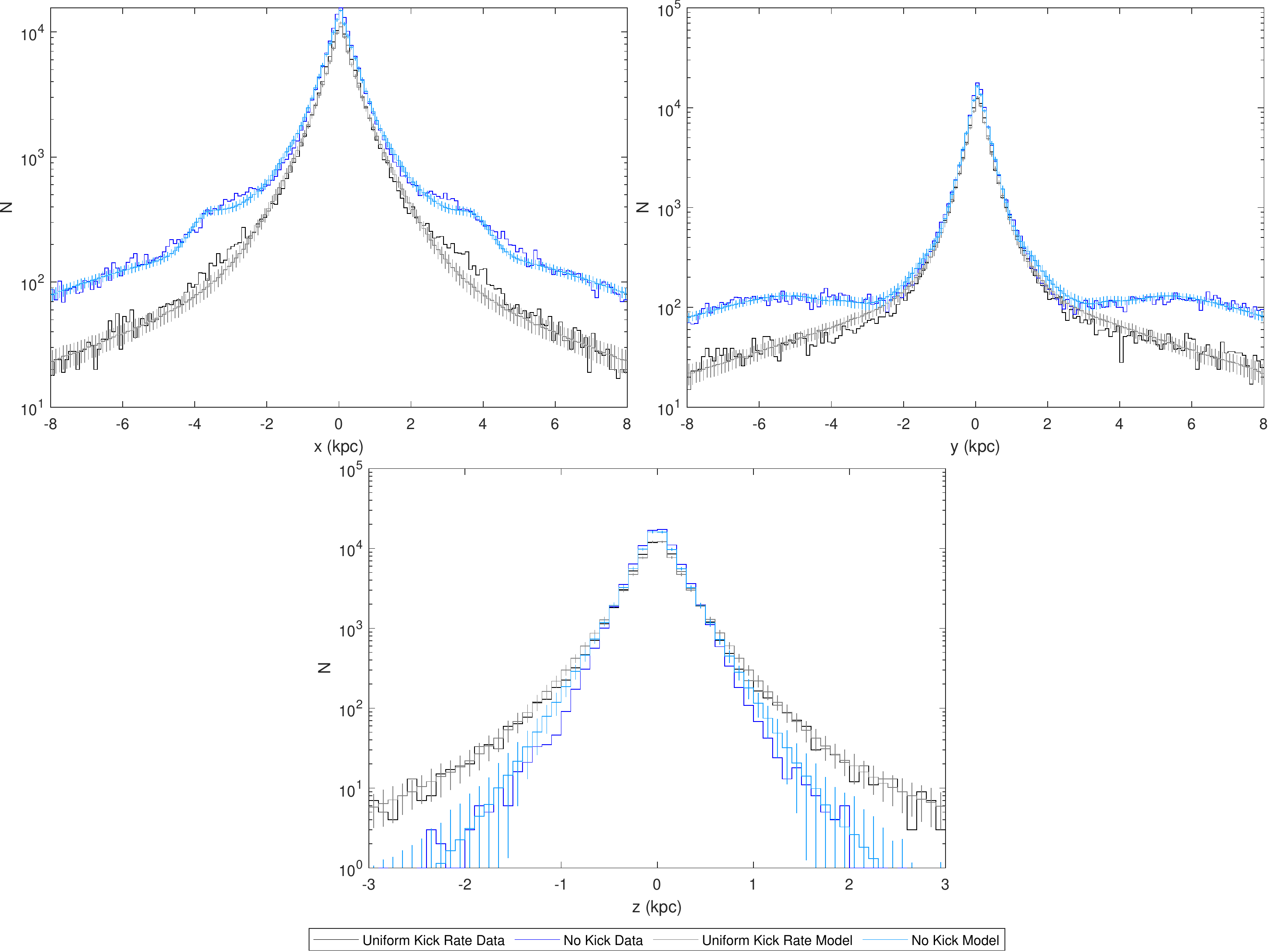}
    \caption{Final profile along $x$, $y$ and $z$ axes with a uniform kick rate generated from the MWb initial conditions. We show both $N$-body simulation data and data simulated using the fitted model. For the fitted model we show the mean number of particles in each bin and the standard deviation. }
    \label{fig:MWb_1d_profiles_uniform_kick_rate}
\end{figure}

\begin{figure}
    \centering
    \includegraphics[width=0.99\linewidth]{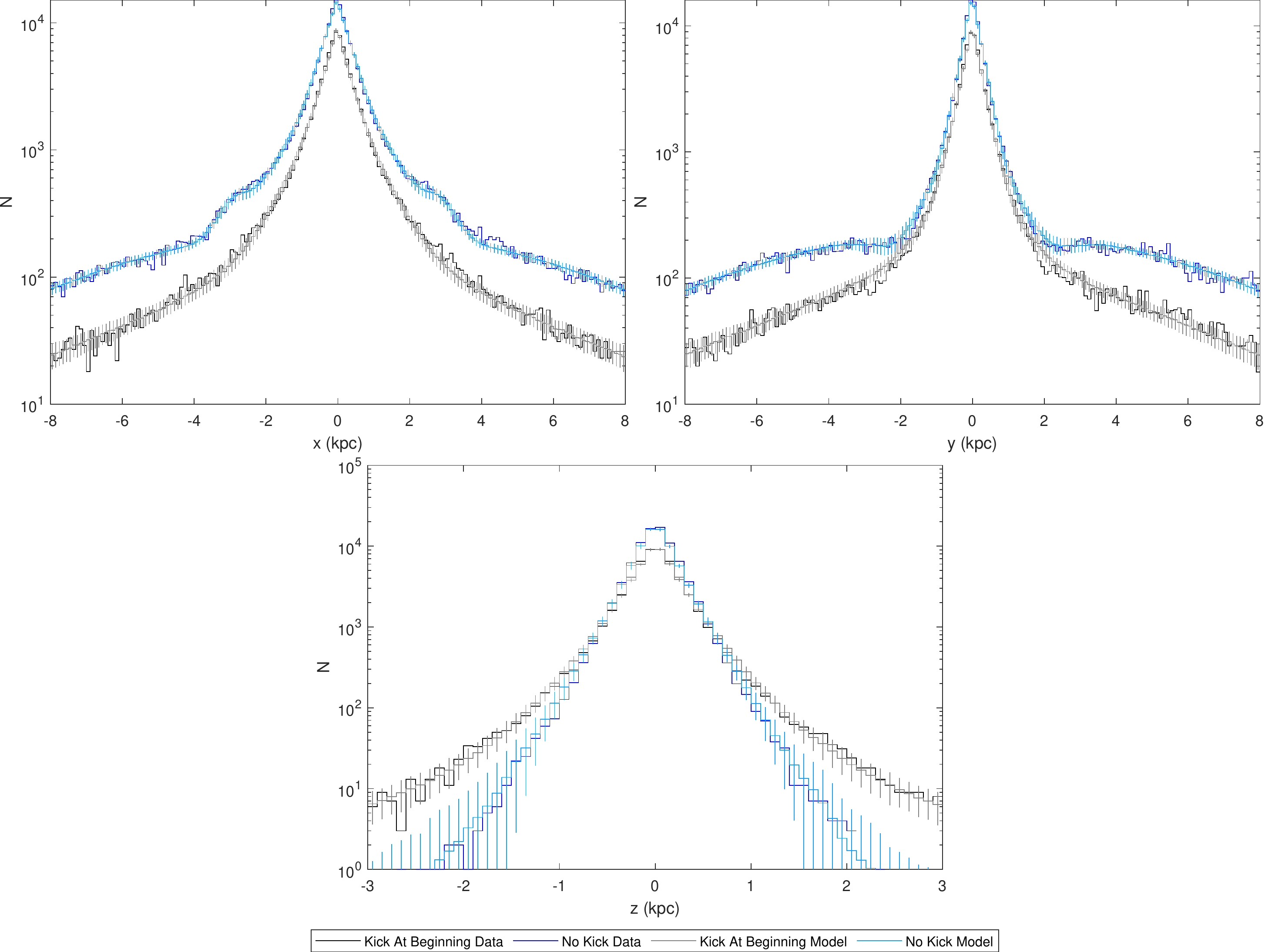}
    \caption{Final profile along $x$, $y$ and $z$ axes with kicks occurring at the beginning generated from the MWc0.8 initial conditions. We show both $N$-body simulation data and data simulated using the fitted model. For the fitted model we show the mean number of particles in each bin and the standard deviation. }
    \label{fig:MWc0.8_1d_profiles_kick_at_beginning}
\end{figure}

\begin{figure}
    \centering
    \includegraphics[width=0.99\linewidth]{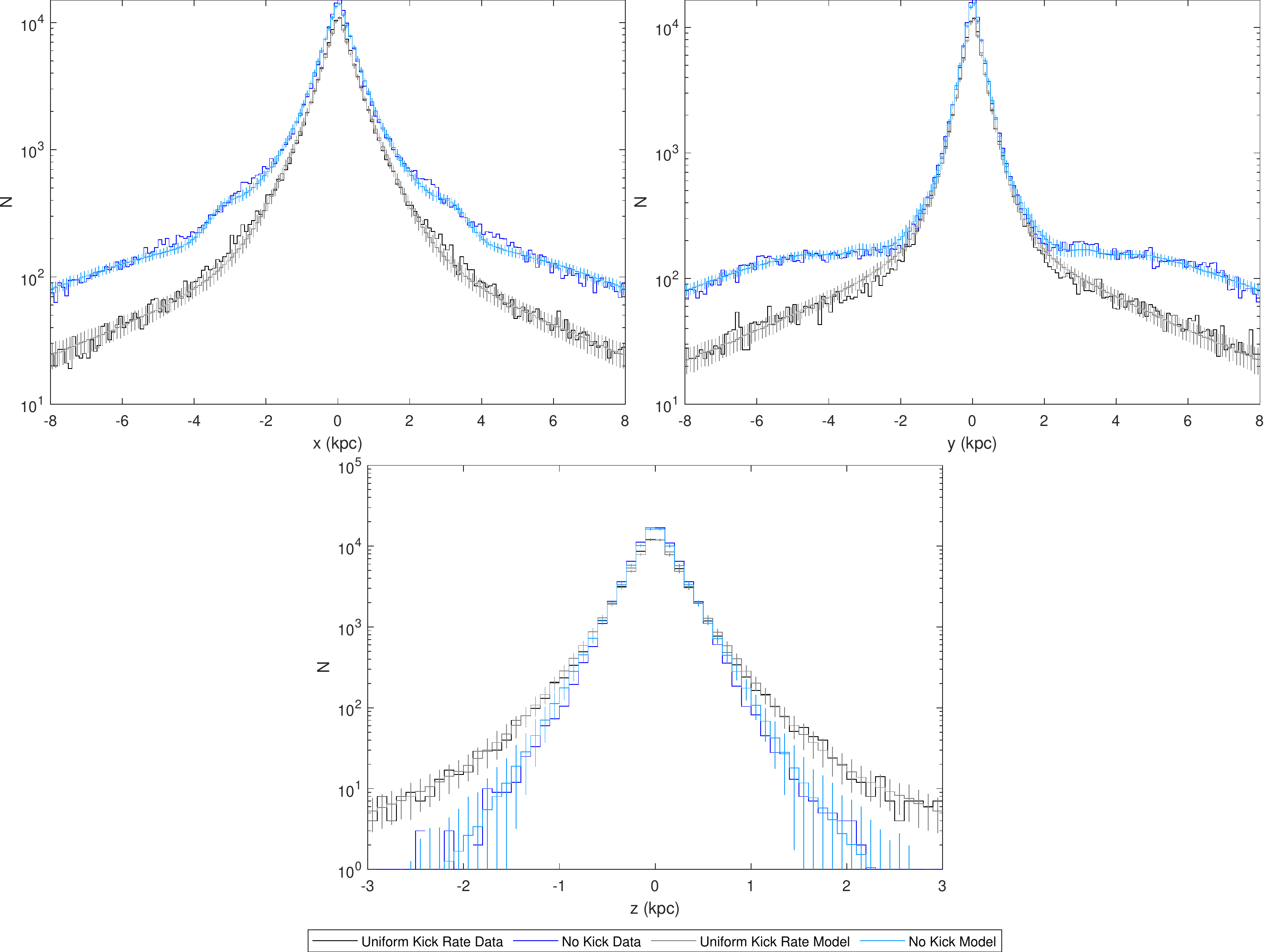}
    \caption{Final profile along $x$, $y$ and $z$ axes with a uniform kick rate generated from the MWc0.8 initial conditions. We show both $N$-body simulation data and data simulated using the fitted model. For the fitted model we show the mean number of particles in each bin and the standard deviation. }
    \label{fig:MWc0.8_1d_profiles_uniform_kick_rate}
\end{figure}

\begin{figure}
    \centering
    \includegraphics[width=0.99\linewidth]{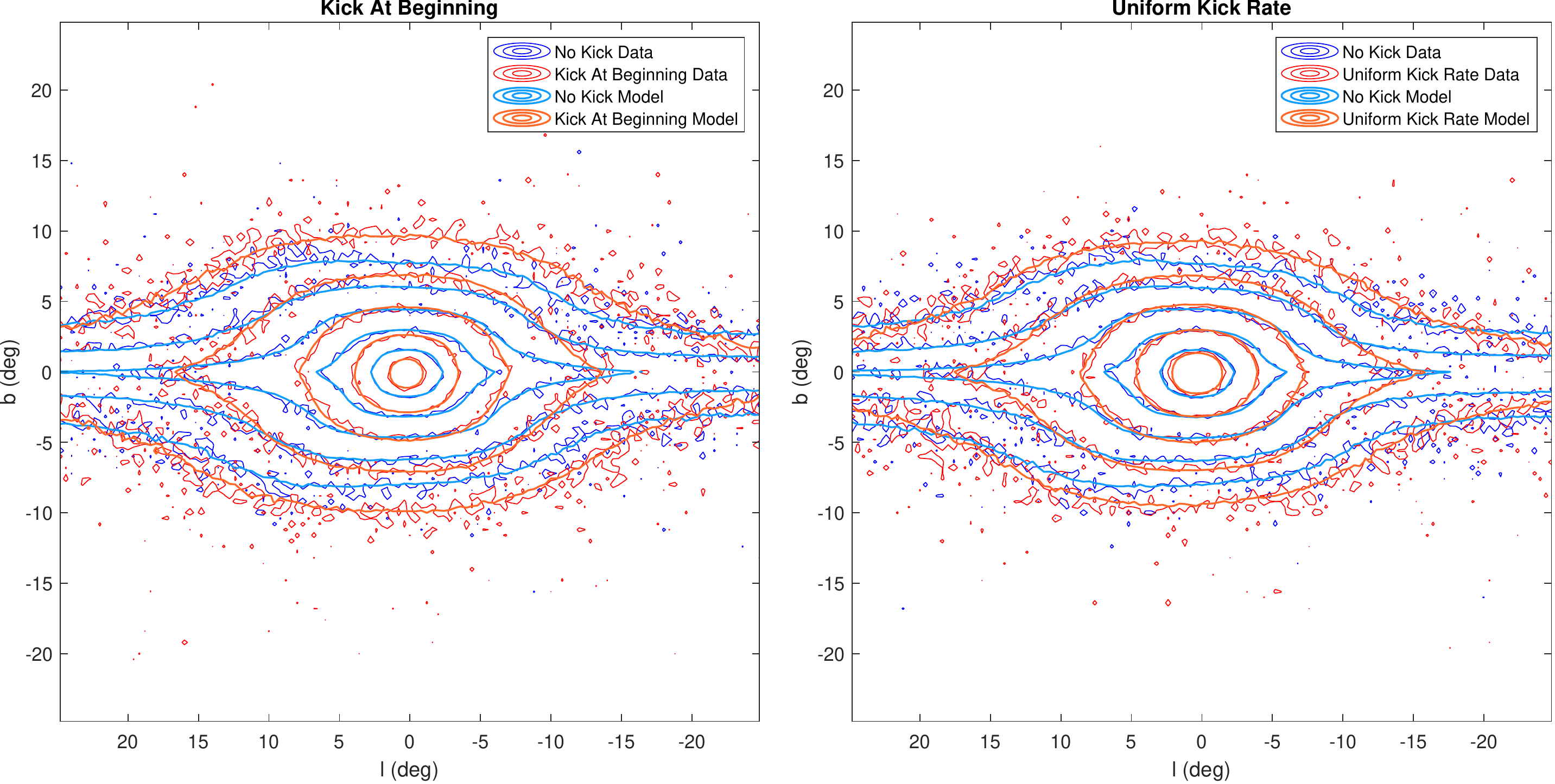}
    \caption{Final flux distribution in Galactic coordinates generated from the MWb initial conditions. The contours for each distribution are at $1$, $2$, $4$, $8$ and $16$ times the mean in this region. The Sun is placed at a distance of $7.9$ kpc, at an angle relative to the bar of $20^\circ$ and at a height of $15$ pc.  }
    \label{fig:MWb_data_los}
\end{figure}

\begin{figure}
    \centering
    \includegraphics[width=0.99\linewidth]{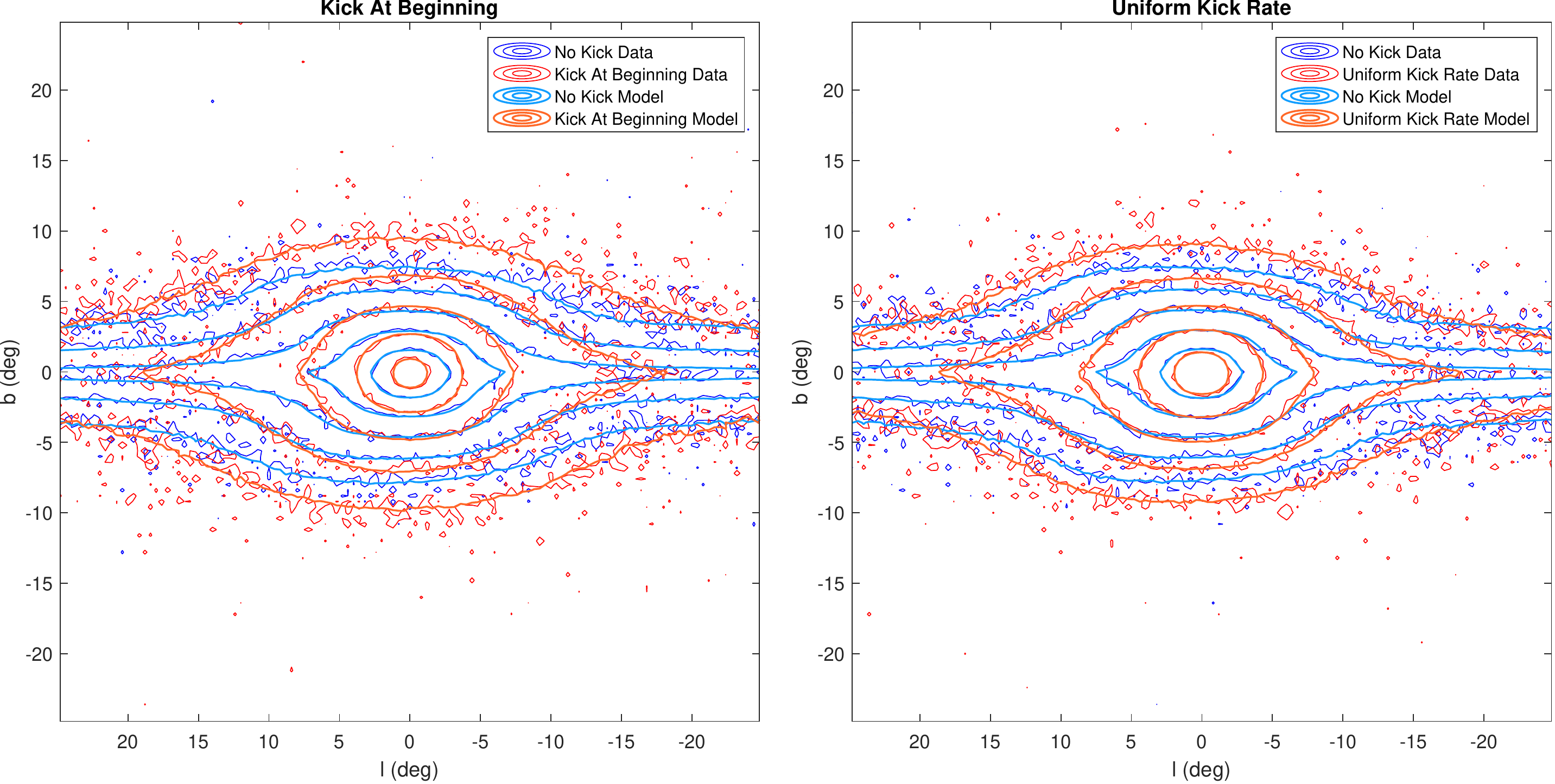}
    \caption{Final flux distribution in Galactic coordinates generated from the MWc0.8 initial conditions. The contours for each distribution are at $1$, $2$, $4$, $8$ and $16$ times the mean in this region. The Sun is placed at a distance of $7.9$ kpc, at an angle relative to the bar of $20^\circ$ and at a height of $15$ pc.  }
    \label{fig:MWc0.8_data_los}
\end{figure}


\begin{figure}
    \centering
    \includegraphics[width=0.99\linewidth]{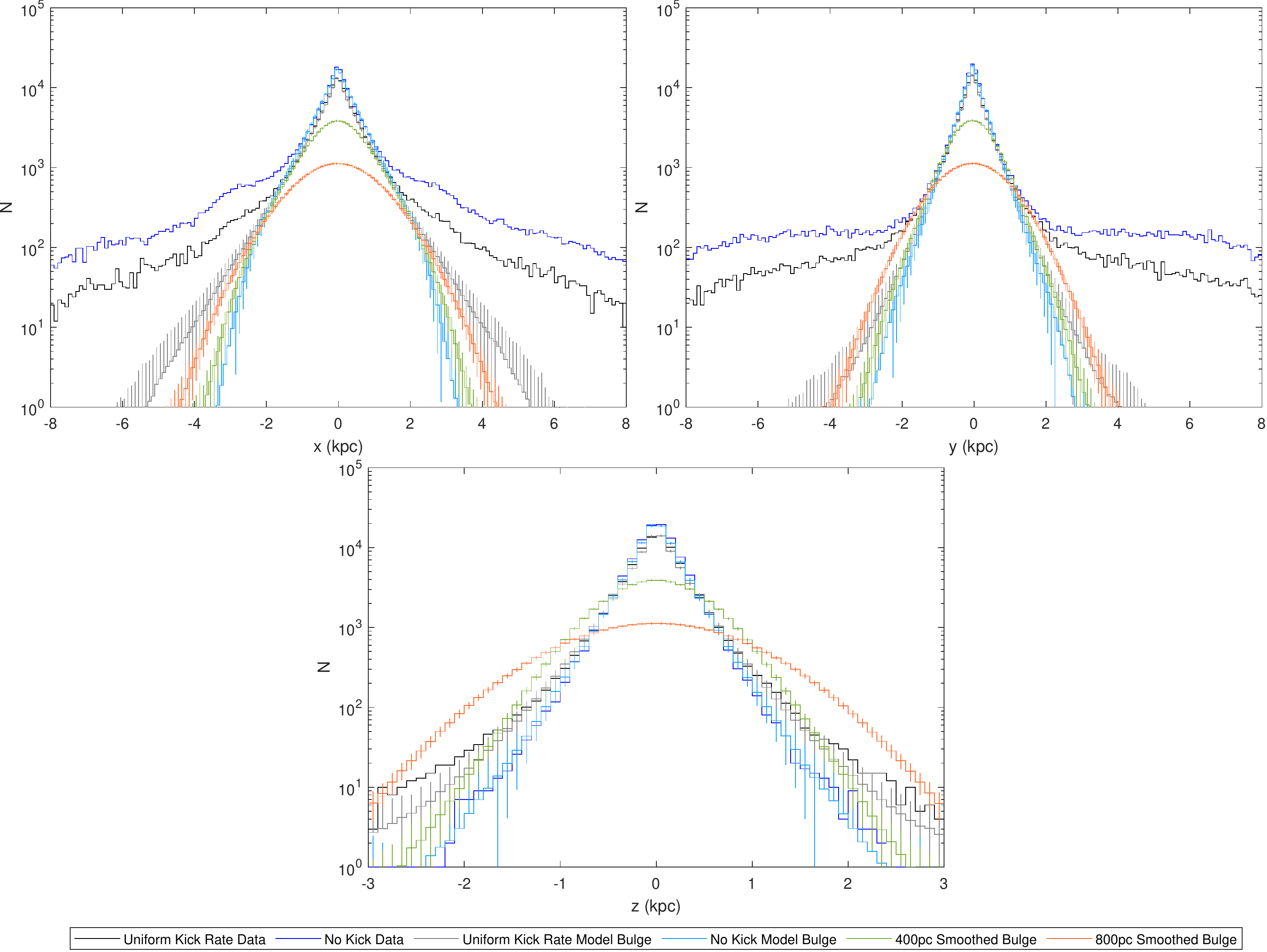}
    \caption{Final profile along $x$, $y$ and $z$ axes with a uniform kick rate generated from the MWa initial conditions. Here we show the fitted bulge components, consisting of the bar plus Hernquist bulge, as well as the no kick bulge smoothed with $400$ pc and $800$ pc Gaussians. We show both $N$-body simulation data and data simulated using the fitted model. For the fitted model we show the mean number of particles in each bin and the standard deviation. }
    \label{fig:MWa_bulge_smoothed_1d_profiles_uniform_kick_rate}
\end{figure}

\begin{figure}
    \centering
    \includegraphics[width=0.99\linewidth]{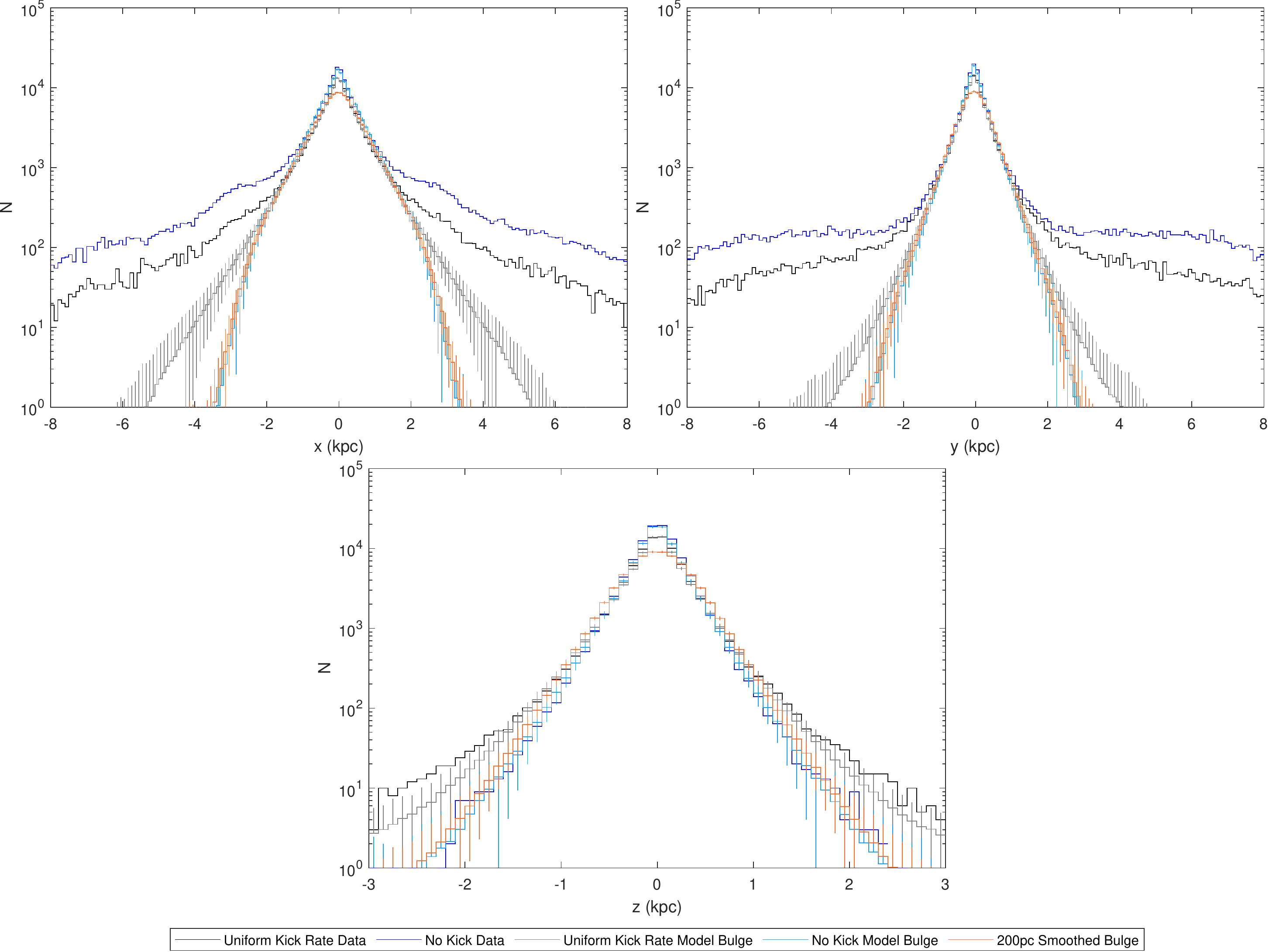}
    \caption{Final profile along $x$, $y$ and $z$ axes with a uniform kick rate generated for the MWa initial conditions. Here we show the fitted bulge components as well as the no kick bulge smoothed with a $200$ pc Gaussian. We show both $N$-body simulation data and data simulated using the fitted model. For the fitted model we show the mean number of particles in each bin and the standard deviation. }
    \label{fig:MWa_bulge_smoothed_200pc_1d_profiles_uniform_kick_rate}
\end{figure}

\begin{figure}
    \centering
    \includegraphics[width=0.99\linewidth]{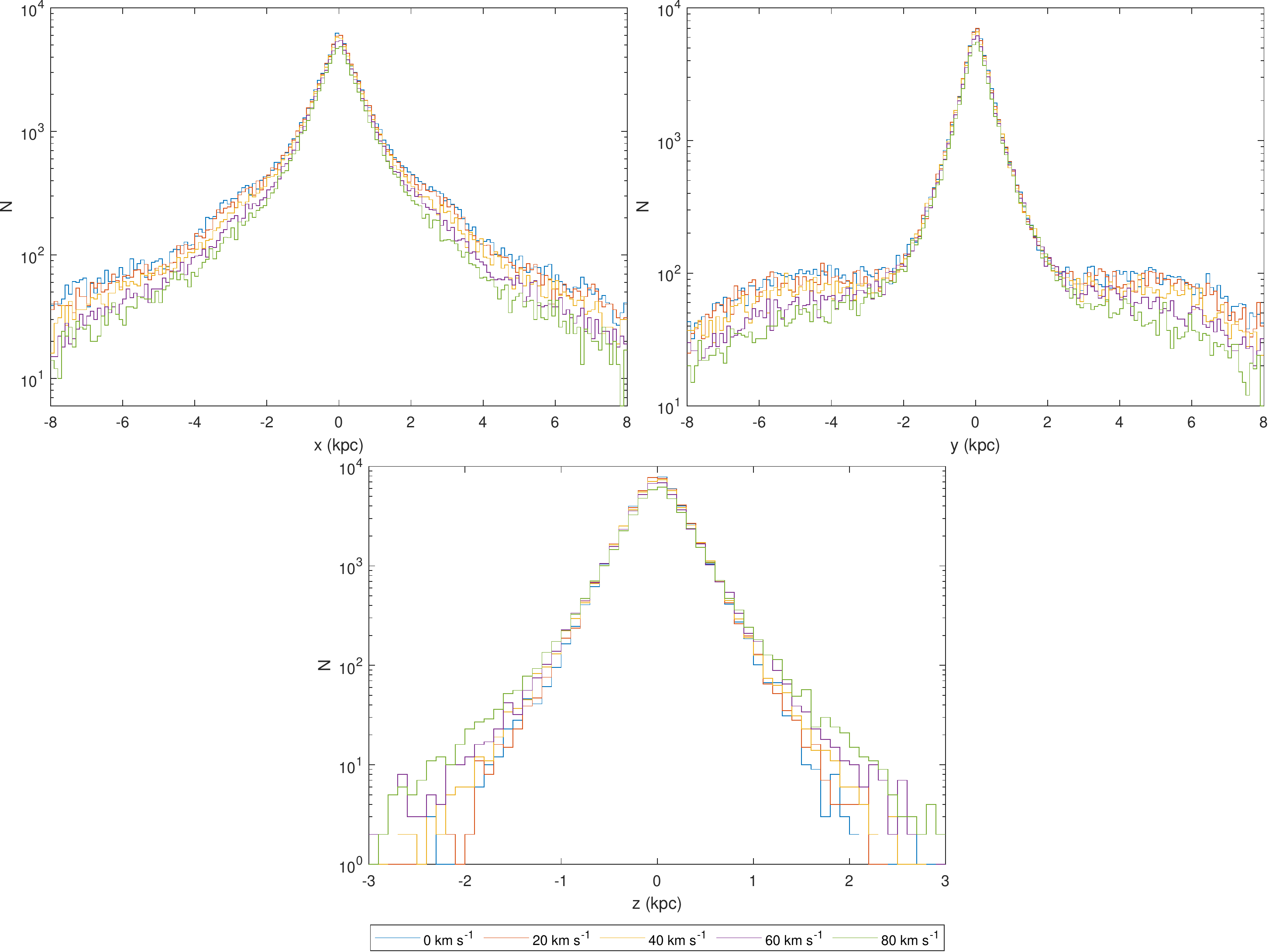}
    \caption{Final profile along $x$, $y$ and $z$ axes with a uniform kick rate for kicks between $0$ km s$^{-1}$ and $80$ km s$^{-1}$ in $20$ km s$^{-1}$ increments. Generated from the MWa initial conditions.}
    \label{fig:MWa_less_80km_per_s_kick_1d_profiles_uniform_kick_rate}
\end{figure}

\section{Markov Chain Monte Carlo Method}
\label{app:mcmc}

In previous work, such as ref.~\cite{Ploeg2020}, we used for our MCMC the adaptive Metropolis algorithm of Haario et al.~\cite{Haario01}. In this article, however, we found it was necessary to replace adaptive Metropolis algorithm with an alternative algorithm to ensure rapid convergence to the peak likelihood region of the parameter space. The MCMC algorithm used in this article is similar to that of Foreman-Mackay et al.~\cite{ForemanMackey2013} with a mixture of the Differential Evolution \cite{TerBraak2006} and snooker updates \cite{TerBraak2008}. Instead of performing a single random walk through the parameter space where proposed moves are accepted with a probability based on the likelihood of the current point of the chain and the proposed next point,
 we use an ensemble of $K$ ``walkers'' where a proposed update for a walker $j$ depends on the distribution of the other walkers.

A single step of the stretch move update suggested in Foreman-Mackay et al.~\cite{ForemanMackey2013} involves updating the $K$ walkers sequentially. Let $\pmb{x}_j$ be the state of walker $j$, an update for walker $\pmb{x}_j$ is performed as follows:
\begin{enumerate}
    \item Draw $k$ from $1,2,...,K$ where $k \neq j$
    \item Draw $z$ from the probability density function with parameter $a$ (Foreman-Mackay et al.~\cite{ForemanMackey2013} suggest $a=2$):
    \begin{equation}
        g(z) \propto \begin{cases}
            \frac{1}{\sqrt{z}} & z \in \left[\frac{1}{a}, a \right] \\
            0 & {\rm otherwise} \\
        \end{cases}
    \end{equation}
    \item Calculate proposal $\pmb{y} = \pmb{x}_k + z \left(\pmb{x}_j - \pmb{x}_k \right)$
    \item Calculate acceptance probability $r$:
    \begin{equation}
        r = z^{d - 1} \frac{p(\pmb{y})}{p(\pmb{x}_j)}
    \end{equation}
    where $d$ is the number of dimensions
    \item Set $\pmb{x}_j = \pmb{y}$ with probability $\min(1,r)$
\end{enumerate}
We found better results using a mixture of two alternative updates: $80\%$ the Differential Evolution update of ter Braak \cite{TerBraak2006} and $20\%$ the snooker update of ter Braak and Vrugt \cite{TerBraak2008}.

To update walker $\pmb{x}_j$ using the Differential Evolution update:
\begin{enumerate}
    \item Draw $k$ and $l$ from $1,2,...,K$ where $k \neq j$, $l \neq j$ and $l \neq k$
    \item Propose $\pmb{y} = \pmb{x}_j + \gamma \left(\pmb{x}_k - \pmb{x}_l \right) + \pmb{e}$ where $\gamma$ is a parameter and where $\pmb{e}$ is drawn from a small $d$ dimensional symmetric probability distribution
    \item Calculate acceptance probability $r$:
    \begin{equation}
        r = \frac{p(\pmb{y})}{p(\pmb{x}_j)}
    \end{equation}
    \item Set $\pmb{x}_j = \pmb{y}$ with probability $\min(1,r)$
\end{enumerate}
We drew $\pmb{e}$ from a $d$ dimensional Gaussian with standard deviation $10^{-5}$ in each dimension. In case the likelihood distribution had multiple modes, we used $\gamma = 1$ with probability $0.1$ as suggested by ter Braak \cite{TerBraak2006}, otherwise we used the default value of $\gamma = 2.38 / \sqrt{2 d}$.

Using the snooker update, we update $\pmb{x}_j$ as follows:
\begin{enumerate}
    \item Draw $k$, $l$ and $m$ from $1,2,...,K$ with no index repeated or equal to $j$
    \item Calculate the orthogonal projections of $\pmb{x}_l$ and $\pmb{x}_m$ onto the line $\pmb{x}_j - \pmb{x}_k$, $\mathrm{proj}_{\pmb{x}_j - \pmb{x}_k}(\pmb{x}_l)$ and $\mathrm{proj}_{\pmb{x}_j - \pmb{x}_k}(\pmb{x}_m)$, where:
    \begin{equation}
        \mathrm{proj}_{\pmb{u}}(\pmb{v}) = \frac{\pmb{v} \cdot \pmb{u}}{\pmb{u} \cdot \pmb{u}} \pmb{u}
    \end{equation}
    \item Propose $\pmb{y} = \pmb{x}_j + \gamma_s \left(\mathrm{proj}_{\pmb{x}_j - \pmb{x}_k}(\pmb{x}_l) - \mathrm{proj}_{\pmb{x}_j - \pmb{x}_k}(\pmb{x}_m) \right)$ where $\gamma_s$ is a parameter
    \item Calculate acceptance probability $r$:
    \begin{equation}
        r = \frac{p(\pmb{y}) \abs{\pmb{y} - \pmb{x}_k}^{d-1}}{p(\pmb{x}_j) \abs{\pmb{x}_j - \pmb{x}_k}^{d-1}}
    \end{equation}
    \item Set $\pmb{x}_j = \pmb{y}$ with probability $\min(1,r)$
\end{enumerate}
We use $\gamma_s = 2.38 / \sqrt{2}$ as suggested by ter Braak and Vrugt \cite{TerBraak2008}.

We also used a simple annealing method in which we divide the log-likelihood by a temperature $T$ which is gradually reduced to $1$. The posterior probability density for a parameter set $\theta$ at MCMC iteration $t$ is:
\begin{equation}
    p(\theta \vert \textrm{$N$-body data}) \propto p(\theta) L^{1/T_t}
\end{equation}
\noindent where $p(\theta)$ is the prior, $L$ is the likelihood, and $T_t$ is the temperature. We used a linearly decreasing $\log(T)$ from $\log(1000)$ to $\log(1)$ during the first half of each Markov chain, which we discard. This method allows the algorithm to explore a broad region in the parameter space, while slowly converging to the desired posterior distribution where $T = 1$. This appeared to help the Markov chains avoid getting stuck in local likelihood maxima.

\end{document}